\documentclass[twocolumn]{aastex63}

\usepackage{epsf}
\usepackage{amsmath}                
\usepackage{amsfonts}               
\usepackage{amssymb}                
\usepackage{epsfig}                 
\usepackage{float}
\usepackage{color}
\usepackage{tabularx}
\usepackage{comment}
\usepackage{makecell}
\usepackage[caption=false]{subfig}
\usepackage{longtable}
\usepackage{array}

\usepackage{hyperref}
\hypersetup{
    colorlinks=true,
    linkcolor=blue,
    filecolor=magenta,      
    urlcolor=cyan,
    pdftitle={Overleaf Example},
    pdfpagemode=FullScreen,
    }
\usepackage{lineno}

\def \cm{~\rm{cm}}
\def \s{~\rm{s}}
\def \km{~\rm{km}}

\def \K{~\rm{K}}

\def \g{~\rm{g}}
\def \G{~\rm{G}}

\begin{document}


\title{Nuclear Neural Networks: Emulating Late Burning Stages in Core Collapse Supernova Progenitors}

\author[0000-0002-2215-1841]{Aldana Grichener}
\affiliation{Steward Steward Observatory and Department of Astronomy, University of Arizona, 933 North Cherry Avenue, Tucson, AZ 85721, USA}
\affiliation{Max Planck Institute for Astrophysics, Karl-Schwarzschild-Str. 1, 85748 Garching, Germany}
\affiliation{Department of Physics, Technion, Haifa, 3200003, Israel}
\correspondingauthor{Aldana Grichener}
\email{agrichener@arizona.edu}

\author[0000-0002-6718-9472]{Mathieu Renzo}
\affiliation{Steward Steward Observatory and Department of Astronomy, University of Arizona, 933 North Cherry Avenue, Tucson, AZ 85721, USA}

\author[0000-0002-0479-7235]{Wolfgang E. Kerzendorf}
\affiliation{Department of Computational Mathematics, Science, and Engineering, Michigan State University, East Lansing, MI 48824, USA}
\affiliation{Department of Physics and Astronomy, Michigan State University, East Lansing, MI 48824, USA}

\author[0000-0003-3441-7624]{Rob Farmer}
\affiliation{Max Planck Institute for Astrophysics, Karl-Schwarzschild-Str. 1, 85748 Garching, Germany}

\author[0000-0001-9336-2825]{Selma E. de Mink}
\affiliation{Max Planck Institute for Astrophysics, Karl-Schwarzschild-Str. 1, 85748 Garching, Germany}
\affiliation{Ludwig-Maximilians-Universitat Munchen, Geschwister-Scholl-Platz 1, 80539 Munchen, Germany}

\author[0000-0003-4456-4863]{Earl Patrick Bellinger}
\affiliation{Department of Astronomy, Yale University, New Haven, CT 06511, USA}

\author[0000-0001-6337-6126]{Chi-kwan Chan}
\affiliation{Steward Observatory and Department of Astronomy, University of Arizona, 933 North Cherry Avenue, Tucson, AZ 85721, USA}
\affiliation{Data Science Institute, University of Arizona, 1230 N. Cherry Avenue, Tucson, AZ 85721, USA}
\affiliation{Program in Applied Mathematics, University of Arizona, 617 North Santa Rita, Tucson, AZ 85721, USA}

\author[0000-0001-7571-0742]{Nutan Chen}
\affiliation{Machine Learning Research Lab, Volkswagen AG, 38440 Munich, Germany}

\author[0000-0002-5794-4286]{Ebraheem Farag}
\affiliation{Department of Astronomy, Yale University, New Haven, CT 06511, USA}

\author[0000-0001-7969-1569]{Stephen Justham}
\affiliation{Max Planck Institute for Astrophysics, Karl-Schwarzschild-Str. 1, 85748 Garching, Germany}

\begin{abstract}
One of the main challenges in modeling massive stars to the onset of core collapse is the computational bottleneck of nucleosynthesis during advanced burning stages. The number of isotopes formed requires solving a large set of fully-coupled stiff ordinary differential equations (ODEs), making the simulations computationally intensive and prone to numerical instability. To overcome this barrier, we design a nuclear neural network (NNN) framework with multiple hidden layers to emulate nucleosynthesis calculations and conduct a proof-of-concept to evaluate its performance. The NNN takes the temperature, density and composition of a burning region as input and predicts the resulting isotopic abundances along with the energy generation and loss rates. We generate training sets for initial conditions corresponding to oxygen core depletion and beyond using large nuclear reaction networks, and compare the predictions of the NNNs to results from a commonly used small net. We find that the NNNs improve the accuracy of the electron fraction by $280-660\:\%$, the average atomic and mass numbers by $150-360 \%$ and the nuclear energy generation by $250-750\:\%$, consistently outperforming the small network across all timesteps. They also achieve significantly better predictions of neutrino losses on relatively short timescales, with improvements ranging from $100-10^{6}\:\%$. While further work is needed to enhance their accuracy and applicability to different stellar conditions, integrating NNN trained models into stellar evolution codes is promising for facilitating large-scale generation of core-collapse supernova (CCSN) progenitors with higher physical fidelity.

\end{abstract}


\section{Introduction}
\label{sec:Intro}

Massive stars ($M\gtrsim 8~\rm{M}_\odot$) usually end their lives in core-collapse supernova (CCSN) events where the core of the star collapses to form a neutron star or a black hole releasing gravitational energy. The explosion mechanism of CCSNe has been widely studied over the past decades (e.g., \citealt{BetheWilson1985}; \citealt{Burrowsetal1995}; \citealt{LangerWoosley1996}; \citealt{Hegeretal2000}; \citealt{Thompsonetal2003}; \citealt{WoosleyJanka2005}; \citealt{Burrowsetal2006}; \citealt{OConnorOtt2011}; \citealt{PapishSoker2011}; \citealt{KushnirKatz2015}; \citealt{Jankaetal2016}; \citealt{Sukhboldetal2016}; \citealt{BlumKushnir2016}; \citealt{Fischeretal2018}; \citealt{Vartanyanetal2019}; \citealt{Ertletal2020}; \citealt{Burrowsetal2020}; \citealt{ShishkinSoker2021}; \citealt{Kresseetal2021};   \citealt{ShishkinSoker2023}; \citealt{Burrowsetal2023}; \citealt{Burrowsetal2024a}; \citealt{Shishkinetal2024}; \citealt{BoccioliRoberti2024}; \citealt{Janka2025}). However, simulations of CCSN still lead to conflicting results. While the physical processes driving these explosions remain debated (e.g., \citealt{Kushnir2015}; \citealt{BurrowsVartanyan2021}; \citealt{Soker2024} and references therein), it is widely agreed upon that the large uncertainties in the evolution and the pre-collapse structure of CCSN progenitors present a major challenge for understanding the CCSN problem and performing multi-dimensional simulations (e.g., \citealt{Farmeretal2016}; \citealt{Sukhboldetal2016}; \citealt{Renzoetal2017}; \citealt{Ottetal2018}; \citealt{Kurodaetal2018}; \citealt{Davisetal2019}; \citealt{Laplaceetal2021}; \citealt{Fields2022}; \citealt{Agrawaletal2022}; \citealt{Renzoetal2024}; \citealt{Burrowsetal2024b}). 

In particular, the large number of isotopes formed in advanced burning stages introduces a significant computational bottleneck. This arises from the stiffness of the extensive set of fully-coupled ordinary differential equations (ODEs) \footnote{A set of ODEs is considered 'stiff' when the ratio of the largest to the smallest absolute eigenvalue of its Jacobian matrix is very large, indicating widely varying scales within the system.} governing stellar evolution and nucleosynthesis, making it extremely challenging to evolve stellar models with large nuclear reaction networks beyond core carbon burning. As a result, studies of late burning stages in massive stars are often performed using small nuclear reaction networks containing around 20 isotopes (e.g., \citealt{AguileraDenaetal2020}; \citealt{FieldsCouch2020}; \citealt{ShishkinSoker2023}; \citealt{Rizzutietal2024}; \citealt{Laplaceetal2024}), known as the `\textit{approx21} family' (modern version based on the idea in \citealt{Timmesetal2000}; see Appendix B in \citealt{Marchantetal2019}). These $\alpha$-chain based nuclear reaction networks are designed to obtain reasonably accurate energy generation rates for most of the stellar lifetime and central electron fractions $Y_{\rm e} \equiv \sum_{i} X_{i}Z_{i} / A_{i}$ (where $X_{i}$, $Z_{i}$ and $A_{i}$ are the isotopes abundances, atomic number and mass number, respectively) similar to large nuclear reaction networks prior to the collapse. 

However, during advanced burning stages, the evolution of the core is highly impacted by weak reactions, which are not included in the small nuclear reaction networks. Performing silicon burning (e.g, \citealt{HixThielemann1996}) and reproducing physical quantities of crucial importance for CCSN explosions, such as the core mass and the mass location of the main nuclear burning, require at least hundreds of isotopes (e.g., \citealt{ThielemannArnet1985}; \citealt{Farmeretal2016}; \citealt{Renzoetal2017}; \citealt{GarciaSenzetal2024}; \citealt{Renzoetal2024}). These will determine the electron fraction that strongly influences the effective Chandrasekhar mass ($M^{\rm eff}_{\rm ch}\propto Y_{\rm e}^{2}$; \citealt{Chandrasekha1931}) and hence the outcome of the core collapse. Moreover, the explosion is sensitive to the structure of the silicon and oxygen shell formed after the end of silicon burning (e.g., \citealt{Ertletal2016}; \citealt{Ottetal2018}; \citealt{Ertletal2020}; \citealt{Wangetal2022}; \citealt{Burrowsetal2023};  \citealt{Burrowsetal2024a}; \citealt{BoccioliFragione2024}) that is dependent on the chosen nuclear reaction network. Computing the pre-supernova (SN) neutrino losses and hence neutrino flux received at Earth requires large nuclear reaction networks that include weak reactions as well  (e.g., \citealt{Pattonetal2017a}; \citealt{Pattonetal2017b};  \citealt{Katoetal2020}).

Therefore, various studies have developed methods to address the computational challenge associated with employing large nuclear reaction networks in stellar nucleosynthesis calculations. These methods include nuclear and quasi statistical equilibrium approximations of isotopes groups (e.g., \citealt{Weaveretal1978}; \citealt{HixThielemann1996}; \citealt{KushnirKatz2020}; \citealt{Zingaleetal2024}; \citealt{Ugolinietal2025}), adaptive nuclear reaction networks (e.g., \citealt{Rauscheretal2002}; \citealt{WoosleyHeger2004}), split burning (e.g., \citealt{Jermynetal2023}), multiplying the reaction rate and energy generation rate by a boosting factor (e.g., \citealt{Cristinietal2017}; \citealt{Georgyetal2024}), and using small networks throughout the simulations combined with large networks for post processing (e.g., \citealt{Sukhboldetal2016}; \citealt{DeanFernandez2024}).

Over the past years, the use of machine learning in theoretical and numerical exploration of nucleosynthesis (e.g., \citealt{Fanetal2022}; \citealt{Zhangetal2025}; see \citealt{SmithLu2024} for a review) and of stellar evolution (e.g., \citealt{Mirouhetal2019}) has increased significantly , with an emphasis on emulating grids of stellar evolution models (e.g, \citealt{Bellingeretal2016}; \citealt{Vermaetal2016}; \citealt{HendriksAerts2019}; \citealt{Bellingeretal2020}; \citealt{Honetal2020}; \citealt{Ksolletal2020}; \citealt{Mombargetal2021};  \citealt{Honetal2024}; \citealt{Maltsevetal2024}; \citealt{Tengetal2024}).

In this study, we employ neural networks, which are universal function approximators (e.g., \citealt{Cybenko1989}), to address the computational challenges posed by nuclear burning in massive stars. We focus on silicon core burning and develop neural network-based emulators, hereafter will be referred to as Nuclear Neural Networks (NNNs), designed to replace the solver of the nucleosynthesis ODEs in stellar evolution and hydrodynamical codes, and carry out a proof-of-concept to assess their performance. NNNs take the logarithm of the temperature and density along with the isotopic composition of a burning region as input, and predict the new composition and energy generation in that region after a given timestep based on the datasets they were trained on.

In Section~\ref{sec:method} we explain how we prepare training sets for our emulators, present their architecture, and describe the training process. In Section~\ref{sec:results} we show the isotopic abundances, electron fraction, nuclear numbers and energy generation predictions of our NNNs and compare their performance to results obtained from nucleosynthesis calculations using a small nuclear reaction network at fixed temperatures and densities. In Section~\ref{sec:applicability} we discuss the applicability of our novel method for stellar evolution and CCSNe simulations. We summarize in Section~\ref{sec:Summary}.

\section{Methods}
\label{sec:method}

We perform nucleosynthesis calculations with large isotope nuclear reaction networks at fixed temperatures and densities (Section~\ref{subsec:NuclearNetwork}) corresponding to advanced burning stages in stars (Section~\ref{subsec:TrainingSets}), and design neural networks (Section~\ref{subsec:Architecture}) that produce machine learning models (Section~\ref{subsec:training}) capable of emulating these results. Integrating these surrogate models into stellar evolution codes will enable the evolution of massive stars with large nuclear reaction networks without the need of solving the nucleosynthesis equations, facilitating the production of reliable CCSN progenitors on a larger scale.  

\subsection{Nuclear Network}
\label{subsec:NuclearNetwork}

We build an emulator for the nuclear reaction network solver of the stellar evolution software instrument Modules for Experiments in Stellar Astrophysics (\textsc{mesa}; \citealt{Paxtonetal2011}; \citealt{Paxtonetal2013}; \citealt{Paxtonetal2015}; \citealt{Paxtonetal2018}; \citealt{Paxtonetal2019}; \citealt{Jermynetal2023}) version r23.05.1. To generate a dataset for training, validation, and testing, we use \textsc{bbq} \citep{bbqFromGitHub}, a wrapper that enables direct access to the solver without modifying its underlying code. \textsc{bbq} uses the Bulrisch-Stoer algorithm \citep{Deuflhard1983} to solve the ODEs of nucleosynthesis 
\begin{equation}
\frac{dX_{i}}{dt} = A_{i} \frac{m_{\rm u}}{\rho}\left(-\sum_j (1+\delta_{ij})r_{ij}(T)+\sum_{k,l} r_{kl,i}(T)\right), 
\label{eq:NucODEs}
\end{equation}
where $X_{i}$ and $A_{i}$ are the abundance and mass of isotope $i$ in the nuclear reaction network, $m_{\rm u}$ is an atomic mass unit, $\rho$ and $\rm T$ are the temperature and density of the burning region, and $r_{ab}$ is the conversion rate of isotopes $a$ to isotope $b$ per unit volume. 

\textsc{bbq} calculates the changes in the composition due to nuclear burning at constant temperatures and densities. Since these calculations are conducted over sufficiently short timesteps, they can be treated independently from the overall stellar evolution. Decoupling the evolution of the nuclear reaction network from the stellar structure equations allows us to follow compositional changes based on the nuclear physics input alone, i.e., without the effect of other physical uncertainties such as mixing induced by convection or rotation. Furthermore, it facilitates the ODE solving process at high temperatures and densities that is extremely challenging otherwise due to the stiffness of the coupled structure and nucleosynthesis equations. 

Using \textsc{bbq} enables obtaining the evolution of compositions that include hundreds of isotopes for a single burning region in a few minutes and therefore to create sufficiently large ($\simeq 10^{6}$) training datasets within $\rm few \times 10^{5}$ CPU hours. Moreover, by leveraging \textsc{bbq}, the problem becomes embarrassingly parallel, as each burning region can be computed independently.

\subsection{Dataset parameter space for emulation}
\label{subsec:TrainingSets}

\begin{figure}
\begin{center}
\vspace*{-1.6cm}
\hspace*{-0.5cm}
\includegraphics[width=0.55\textwidth]{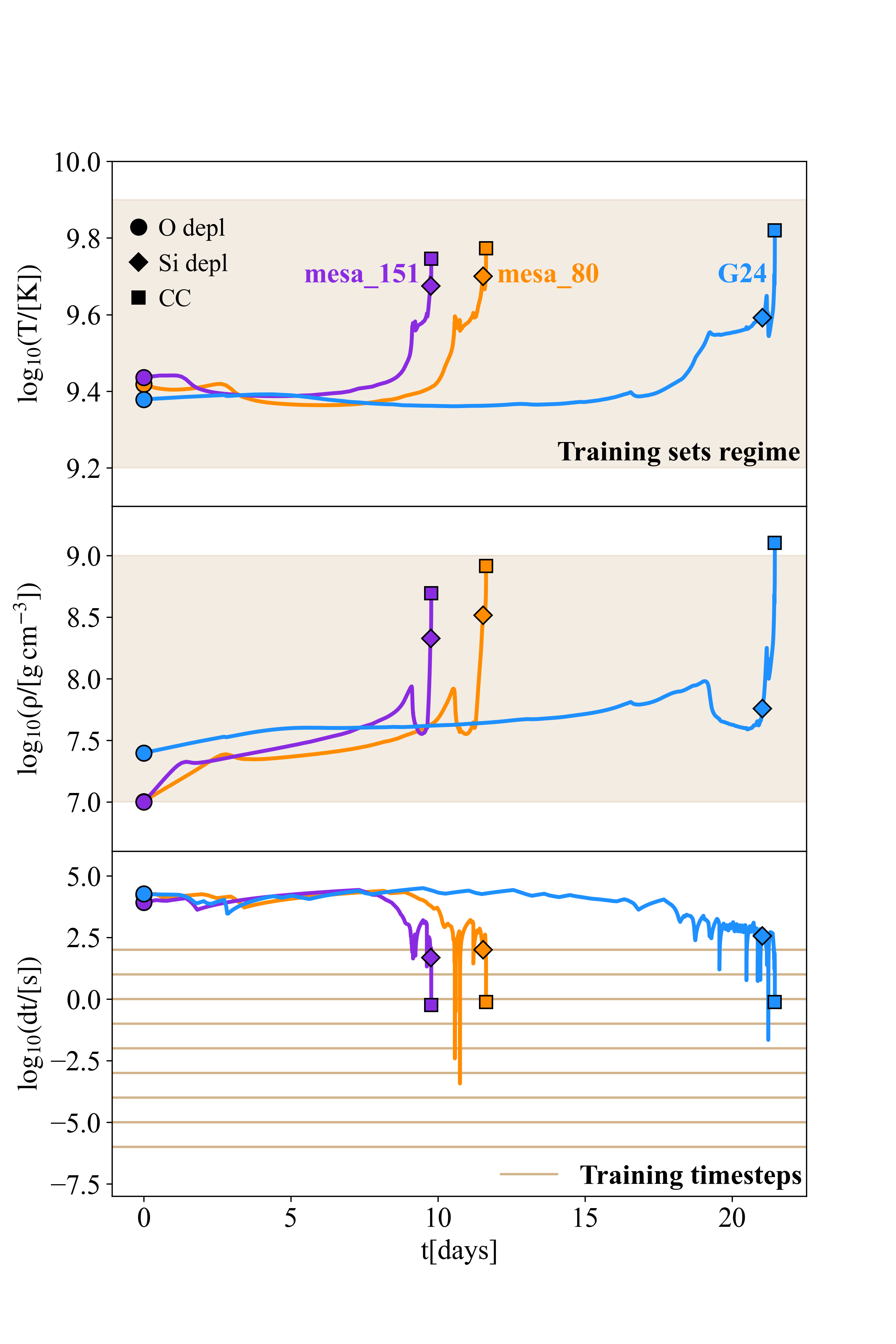}
\vspace*{-1cm}
\caption{For training the nuclear neural networks (NNNs), we consider a range of parameters that correspond to the conditions encountered in stellar models. We show the time evolution of the central temperature (top panel), central density (middle panel) and timesteps taken by simulations (lower panel) of supernova (SN) progenitors with initial masses of $M_{\rm ZAMS} = 20~\rm{M}_\odot$ and a nuclear reaction networks composed of 80 isotopes (orange; this paper) and 151 isotopes (purple; this paper), from post oxygen core burning to collapse.
The light blue curves present the same tracks for a $M_{\rm ZAMS} = 40~\rm{M}_\odot$ chemically homogeneous black hole progenitor taken from \citealt{Gottliebetal2024} (G24). We mark the depletion of oxygen, silicon and the onset of core collapse with circles, diamonds and squares, respectively.  The shaded brown areas (top two panels) are the temperature and density regimes of the training sets, and the brown lines (bottom panel) indicate the timesteps on which separate NNNs were trained. }
\label{fig:ParamasSelection}
\end{center}
\end{figure}

We focus on late burning stages in SN progenitors. To determine the relevant parameter space for emulation, we evolve massive stars with initial masses of $M_{\rm ZAMS}=20~\rm{M}_\odot$ according to the test case \textit{20M\textunderscore pre\textunderscore ms\textunderscore to\textunderscore core\textunderscore collapse} of  \textsc{mesa-r23.05.1} \footnote{We note that the \textsc{mesa} test cases are designed to run rapidly by sacrificing physical and numerical accuracy. Nevertheless, they are sufficient to have an estimate of the range of temperatures and densities during a particular burning phase.}, changing the nuclear reaction network. In Fig.~\ref{fig:ParamasSelection} we present the temperature (top panel) and density (middle panel) tracks of stellar models evolved with the nuclear reaction networks \textit{mesa\textunderscore 80} (orange curves) and \textit{mesa\textunderscore 151} (purple curves), and a rotationally-induced chemically homogeneous evolution of a $M_{\rm ZAMS}=40~\rm{M}_\odot$ black hole progenitor modeled using a nuclear reaction network of 128 isotopes (light blue curves; from \citealt{Gottliebetal2024}). In all cases $t=0$ represents oxygen depletion in the core (central oxygen abundance drops under $10^{-3}$; marked with a circle), and the evolution goes through silicon core burning (silicon depletion occurs when the central silicon abundance drops under $10^{-3}$; marked with a diamond) and continues until the onset of core collapse (infall velocity larger than $300 \km \s ^{-1}$;  marked with a square). We note that while corresponding to different times, the temperature and density regimes of all three models are roughly within the shaded brown ares, which we select as the parameter space for NNN training. This regime is also approximately coincides with the silicon burning study conducted by \cite{HixThielemann1996}. 

We create datasets for training with \textsc{bbq} using two different nuclear reaction networks, \textit{mesa\textunderscore 80} and \textit{mesa\textunderscore 151} (see Appendix \ref{sec:AppendixIsos}), which are determined based on \citealt{Farmeretal2016} and by examining the compositions resulting from networks with a different number of isotopes (see Appendix \ref{sec:AppendixMesaBug}). \textsc{bbq} requires the initial temperature, density and composition of a burning region, and computes the change in its abundances and resultant energy generation and loss rates. For each nuclear reaction network, we construct a grid of temperatures, densities, and electron fractions using the Sobol sampler for quasi-random distribution (\citealt{Sobol1967}; see also Fig.~14 of \citealt{Bellingeretal2016}) to achieve a more uniform coverage of the multi-dimensional space compared to random sampling, and sample the temperatures and densities of each burning region logarithmically in the regimes ${10^{9.2} \K <T<10^{9.9} \K}$ and ${10^{7} \g \cm ^{-3}<\rho < 10^{9} \g \cm ^{-3}}$ (shaded  brown regions of Fig.~\ref{fig:ParamasSelection}). Compositions, however, consist of non-independent abundances that sum to unity, making quasi random sampling non-applicable. Therefore,  for each point in our grid we generate a random initial composition that corresponds to its electron fraction. The electron fraction range, $0.45<Y_{\rm e}<0.5$, was chosen to correspond to values typical of silicon core burning. We note that this sampling method necessarily generates training sets that include initial conditions not produced by stellar evolution models. However, we find that narrowing our parameter space by using typical composition from inner regions of massive star evolved with \textsc{mesa} results in NNNs with poor performance (see Section \ref{sec:applicability} for more details), as the broader coverage significantly enhance the generalizations ability of the NNN trained models.   

Using the temperatures, densities and compositions sampled by the methods above, we generate $1048576 \simeq 10^{6}$ different combinations of temperatures, densities and nuclear compositions for each nuclear network. The dataset size was chosen based on experiments showing that further increasing the training dataset does not improve the loss values or the predictions for the parameters of interest (equations \ref{eq:DeltaComp} and \ref{eq:DeltaYe} - \ref{eq:DeltaEpsNu}). The model likely has reached a saturation point in its learning capacity with respect to the available datasets. 

Each combination is then given as an initial condition to \textsc{bbq}, which evolves the abundance of isotopes in time, creating the compositions we use for training the emulators presented in Section~\ref{subsec:Architecture} using the procedure described in Section~\ref{subsec:training}. The training process is performed with compositions found after 9 logarithmically spaced times from $10^{-6} \s$ to $100 \s$, encompassing most timesteps taken by the \textsc{mesa} simulations in the three aforementioned stellar models (bottom panel of Fig.~\ref{fig:ParamasSelection}; brown dashed lines). We also include timesteps smaller by several orders of magnitude, which might be needed for hydrodynamical simulations. We train a separate neural network for every timestep. The size of the training sets for each NNN is about 3 GB in the case of $mesa\textunderscore80$ and 6 GB for $mesa\textunderscore151$, amounting to approximately 80 GB in total.

\subsection{Nuclear Neural Networks Architecture}
\label{subsec:Architecture}

\begin{figure*}
\begin{center}
\includegraphics[width=0.75\textwidth]{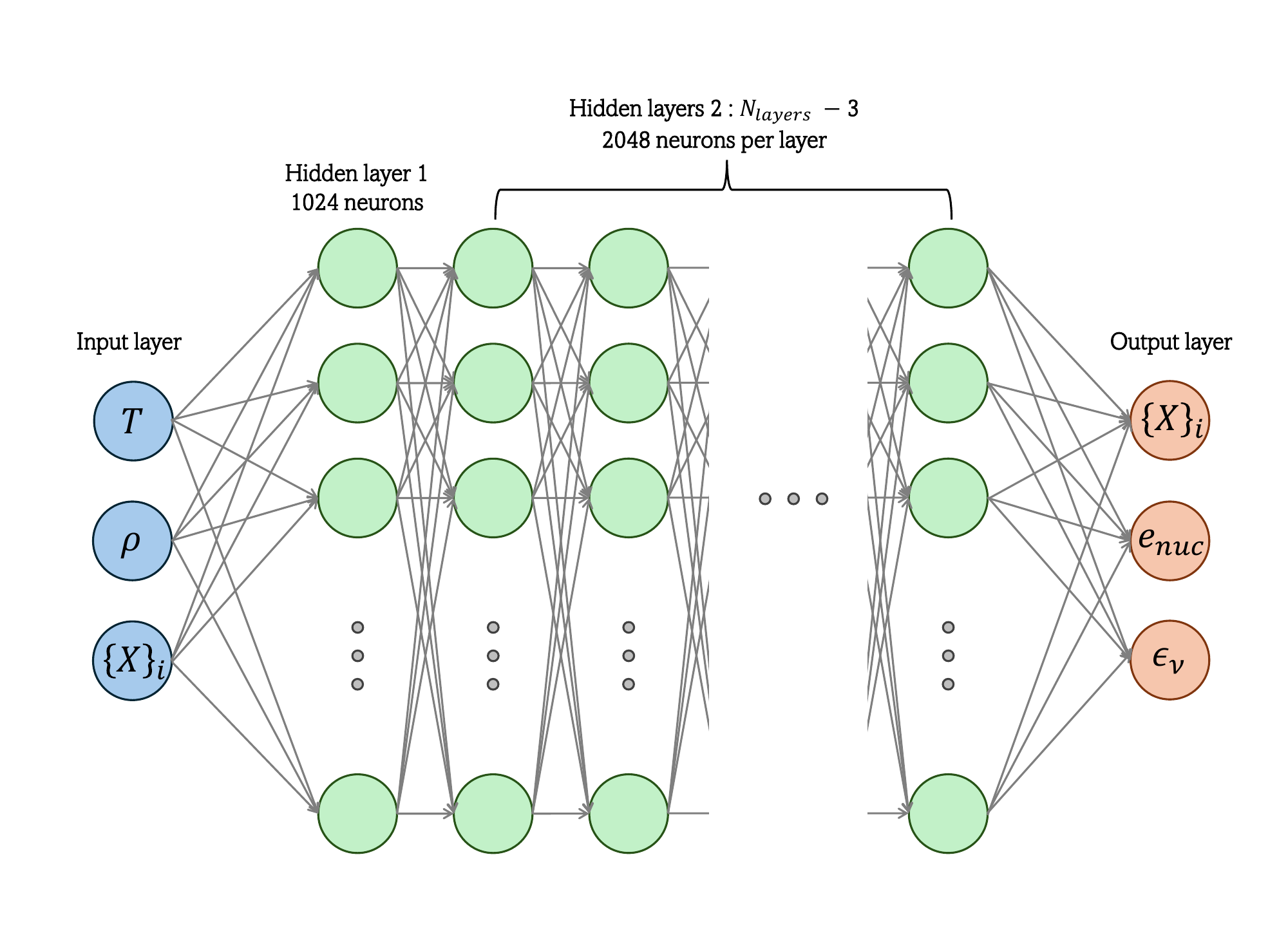}
\caption{Nuclear Neural Networks (NNNs) architecture. The emulators are composed of an input layer that contains the temperature $T$ (on a logarithmic scale), density $\rho$ (on a logarithmic scale) and initial composition $\{X\}_{i}$, several hidden layers and an output layer that consists of the composition, normalized nuclear energy generation term $e_{\rm nuc}$ and normalized neutrino-loss term $\epsilon_{\rm \nu}$ after a given timestep. The input layer maps the initial conditions to a hidden layer of 1024 neurons, and from there they are mapped between the rest of the hidden layers with 2048 neurons each, until we obtain the composition and normalized energy terms predicted by the NNNs. For NNNs trained on datasets produced with the nuclear reaction network \textit{mesa\textunderscore 80} and \textit{mesa\textunderscore 151}, the total number of layers is $N_{\rm layers}=12$ and $N_{\rm layers}=9$, respectively. Each component of the architecture, including the number of layers and number of neutrons per layer, was determined by hyper-parameter search.}
\label{fig:NNNarchitecture}
\end{center}
\end{figure*}


We use \textsc{PyTorch Lightning} \citep{FalconBorovec2020} version 2.4.0, a wrapper of  \textsc{PyTorch} \citep{Paszkeetal2019}, to develop a fully-connected neural network emulator (see Fig.~\ref{fig:NNNarchitecture}). This architecture was chosen for its ability to approximate high-dimensional functions efficiently. It takes as input the logarithm of the temperature $T$ (in $\mathrm{K}$), logarithm of the density $\rho$ (in $\mathrm{g \:}\mathrm{cm}^{-3}$), and nuclear composition $\{X\}_{i}$ of a stellar burning region, and predicts the composition $\{X\}_{i}$, nuclear energy generation per unit mass $e_{\rm nuc}$ (in  $\mathrm{erg}\;\mathrm{g}^{-1}$) and energy lost by neutrinos from weak nuclear reactions per unit mass per unit time $\epsilon_{\rm \nu}$ (in $\mathrm{erg}\;\mathrm{g}^{-1}\;\mathrm{s}^{-1}$) in that burning region after a given timestep $dt$.
The last two outputs satisfy 
\begin{equation}
q = e_{\rm nuc} + \int _0 ^{dt} \epsilon_{\rm \nu} dt' = \int _0 ^{dt} \epsilon_{\rm nuc} dt' + \int _0 ^{dt} \epsilon_{\rm \nu} dt', 
\label{eq:EpsNucEpsNu}
\end{equation}
where $q$ is the sum over the rest mass difference between reactants and products for all nuclear reactions per unit mass and $\epsilon_{\rm nuc}$ is the nuclear energy generation per unit mass per unit time. For simplicity, we will hereafter refer to $e_{\rm nuc}$ as the nuclear energy generation term and $\epsilon_{\rm \nu}$ as the neutrino-loss term.

We normalized the energy terms dividing by $10^{16}$, i.e., ${e_{\rm nuc} \rightarrow e_{\rm nuc}/(10^{16}\; \mathrm{erg}\;\mathrm{g}^{-1})}$ and ${\epsilon_{\rm \nu} \rightarrow \epsilon_{\rm \nu}/(10^{16}\;\mathrm{erg}\;\mathrm{g}^{-1}\;\mathrm{s}^{-1})}$, to facilitate the training by aligning their values more closely with typical composition values and reducing the range of orders of magnitude involved. This normalization fraction gave the best results for all our explored timesteps, and was determined through a systematic study.

The NNNs map the $2+N_{iso}$ inputs, where $N_{iso}$ is the number of isotopes in the nuclear reaction network, into a higher-dimensional representation, enabling the model to extract and identify underlying patterns. The first hidden layer transforms the input to 1024 neurons, followed by subsequent layers that map to 2048 neurons. We find that this gradual expansion to a higher-dimensional representation results in better performance. Each component of the architecture was selected through a manual hyper-parameter search aimed at minimizing errors while balancing performance and computational efficiency. The number of neurons per layer was chosen by testing values between 256 and 4096. The optimal number of layers $N_{\rm layers}$ varies between the two nuclear reaction networks: for NNNs with $N_{\rm iso}=80$ isotopes we find $N_{\rm layers}=12$ results in the best performance, whereas for $N_{\rm iso}=151$ we find $N_{\rm layers}=9$ is the most effective (see Fig.~\ref{fig:LossesDifferentLayers} in Appendix \ref{sec:AppendixNNNtries}).

Each hidden layer is followed by a Rectified Linear Unit (ReLU) activation function \citep{Glorotetal2011}, which introduces non-linearity. ReLU is particularly useful for this task as it helps prevent vanishing gradients, making it easier for the model to learn complex relationships in deep networks. Replacing the activation function of our network with sigmoid or hyperbolic tangent (tanh) yields less accurate results. We fix the size of the output layer at $N_{iso}+2$, and apply the log-softmax activation function to the composition output to ensure the abundances sum to one. The log-scale is critical for predicting very low abundances that might lead to new reactions pathways, and span over many orders of magnitude. We use the same architecture to train NNNs on each of our chosen timesteps.

\subsection{Training Process}
\label{subsec:training}

We follow a train and validation split of our dataset for emulator training \citep[see e.g., ][]{Kerzendorfetal2022} allocating most of the data for training the model. Similar to \citet{Kerzendorfetal2021}, we forgo a traditional test but validate the NNN's prediction abilities by comparing its emulation results on 700 datasets to the current stellar evolution simulations setting (see Section~\ref{sec:results}). We particularly emphasize the composition in the prediction of our validation set and apply early stopping when the composition starts to overfit (see Figure~\ref{fig:LossesVSepoch} in Appendix \ref{sec:AppendixNNNtries} for more details). 

To train the model, we chose a batch size of 512, which has been found to be optimal for this problem with manual hyper-parameter search. We use the L1 loss function \citep{Hermathena1887,Tibshirani1996} on a logarithmic scale for the composition and on a linear scale for the normalized energy terms.  Our choice of the loss function is motivated by the goal of accurately predicting small abundance values on a wide range of scales, as common for nuclear compositions. Using the Huber loss function \citep{Huber1964,Hastieetal2004}, designed to decrease the sensitivity to outliers in the datasets, results in much less accurate predictions (see Appendix \ref{sec:AppendixNNNtries} for an exploration of different choices). We use the Adam optimization algorithm \citep{Kingma2014AdamAM} which is a stochastic gradient descent method for its adaptive learning rate properties. We chose a learning rate of $10^{-4}$ based on preliminary experimentation, ensuring steady convergence without drastic fluctuations. All the networks parameters are selected once the predictions for the composition begin to worsen.
 
\section{Results}
\label{sec:results}

In this Section, we present the predictions of the NNNs, and evaluate the errors in key quantities critical for late burning stages of massive stars and the generation of reliable core-collapase progenitors. We compare these uncertainties to those that result from using the small nuclear reaction network \textit{approx21\textunderscore cr60\textunderscore plus\textunderscore co56}\footnote{We use \textit{approx21\textunderscore cr60\textunderscore plus\textunderscore co56} for the comparison, which is the standard \textit{approx21} nuclear reaction network with the addition of $^{56}\mathrm{Co}$ and the replacement of $^{56}\mathrm{Cr}$ with $^{60}\mathrm{Cr}$, since it was designed to better match the central $Y_{\rm e}$ values produced by large nuclear reaction networks.} (see Appendix \ref{sec:AppendixIsos}), which includes 22 isotopes. We scrutinize the NNNs performance on datasets comprising of approximately ${N_{\rm test} \simeq 700}$ \textsc{bbq} outputs that constitute the test set. These outputs were generated following the same procedure described in Section~\ref{subsec:TrainingSets}, limiting the initial compositions to start with isotopes present in \textit{approx21\textunderscore cr60\textunderscore plus\textunderscore co56} to enable this comparison. While this test does not cover the entire composition parameter space on which the NNNs were trained, performing a similar test without this constraint on the initial compositions produced similar results for the NNN prediction errors.

\begin{figure*}
\begin{center}
\hspace*{-0.45cm}
\includegraphics[width=1.05\textwidth]{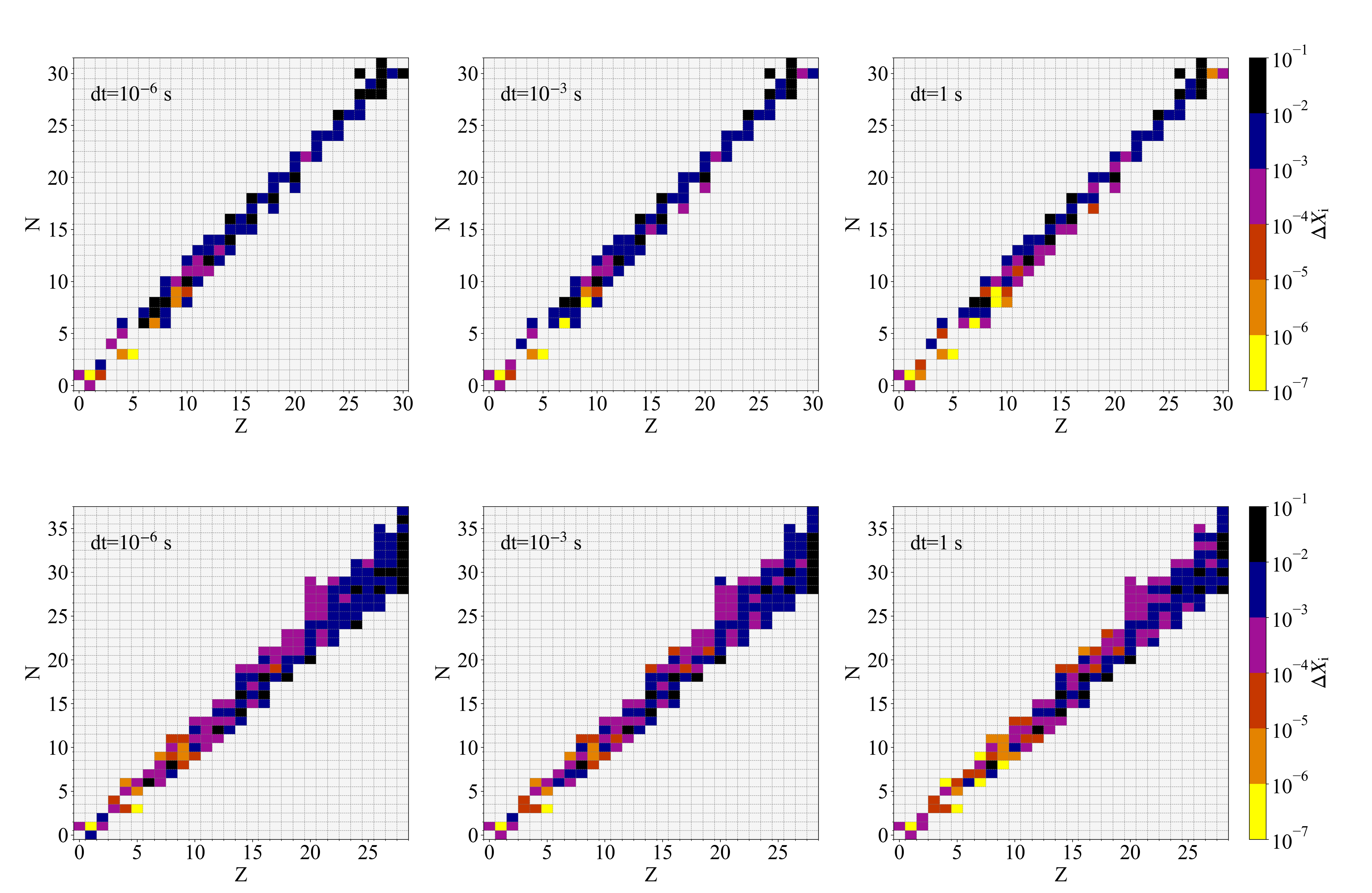}
\caption{Errors in the nuclear neural network (NNN) predictions of the composition for different timesteps. Z is the number of protons of each isotope and N is its number of neutrons. Top row: errors in abundances for NNNs trained on datasets generated using the nuclear reaction network \textit{mesa\textunderscore 80} with a timestep $\rm dt = 10^{-6} \s$ (left panel), $\rm dt = 10^{-3} \s$ (middle panel) and $\rm dt = 1 \s$ (right panel). Bottom row: similar to the top row for NNNs trained on datasets generated using the nuclear reaction network \textit{mesa\textunderscore 151}.
}
\label{fig:AbundanceLossDensityMap}
\end{center}
\end{figure*}

We define the error in the predicted abundance of each isotope compared to the target abundance found by running \textsc{bbq} with a large nuclear reaction network (\textit{mesa\textunderscore 80} or \textit{mesa\textunderscore 151}) as 
\begin{equation}
\delta X_{i} = |X_{i,\rm NNN} - X_{i,\rm target}|, 
\label{eq:deltaComp}
\end{equation}
where $X_{i,\rm NNN}$ is the abundance of an isotope predicted by the NNN and $X_{i,\rm target}$ is the target value the NNN is trying to predict, and average over the errors in our entire test dataset
\begin{equation}
\Delta X_{i} =  \langle \delta X_{i}\rangle = 
\frac{1}{N_{\rm test}}\sum_{j=1}^{N_{\rm test}} | X_{i,\mathrm{NNN},j} - X_{i, \mathrm{target},j}|. 
\label{eq:DeltaComp}
\end{equation}
Fig.~\ref{fig:AbundanceLossDensityMap} shows the average errors in the NNN predictions of the isotopic abundances $\{X\}_{i}$ computed according to equation \ref{eq:DeltaComp} for different timesteps on which the NNNs were trained. We can see that the errors for most isotopes, regardless of the choice of timestep ($dt=10^{-6} \s$, $dt=10^{-3} \s$ or $dt=1 \s$, from left to right) or nuclear reaction network (\textit{mesa\textunderscore 80}; top row, or \textit{mesa\textunderscore 151}; bottom row) fall in the regime of $10^{-4}\lesssim \Delta X_{i} \lesssim 10^{-1}$. Relatively light isotopes have the smallest errors. We chose to show the absolute errors to emphasize the importance of the NNN mispredictions, which grow with the abundance of the isotopes.   

Using our predicted abundances, we compute the error in the electron fraction
\begin{equation}
\delta Y_{\rm e} = \sum_{i=1}^{N_{\rm iso}} \frac{Z_{i}}{A_{i}}|X_{i,\rm NNN}-X_{ i,\rm target}|.
\label{eq:deltaYe}
\end{equation}
Averaging over our test sample we obtain
\begin{equation}
\Delta Y_{\rm e} = \frac{1}{N_{\rm test}} \sum_{j=1}^{N_{\rm test}} \sum_{i=1}^{N_{\rm iso}}  \frac{Z_{i}}{A_{i}}|X_{i,\mathrm{NNN},j}-X_{i,\mathrm{target},j}|.
\label{eq:DeltaYe}
\end{equation}
We apply the same definition of the error to the average mass number $\bar{A}=\sum_i X_iA_i$
\begin{equation}
\Delta \bar{A} = \frac{1}{N_{\rm test}} \sum_{j=1}^{N_{\rm test}} \sum_{i=1}^{N_{\rm iso}}  A_{i}|X_{i,\mathrm{NNN},j}-X_{i,\mathrm{target},j}|,
\label{eq:DeltaAbar}
\end{equation}
average atomic number $\bar{Z}=\sum_i X_iZ_i$
\begin{equation}
\Delta \bar{Z} = \frac{1}{N_{\rm test}} \sum_{j=1}^{N_{\rm test}} \sum_{i=1}^{N_{\rm iso}}  Z_{i}|X_{i,\mathrm{NNN},j}-X_{i,\mathrm{target},j}|,
\label{eq:DeltaZbar}
\end{equation}
energy generation term
\begin{equation}
\Delta e_{\rm nuc} = \frac{1}{N_{\rm test}} \sum_{j=1}^{N_{\rm test}} |e_{\rm nuc,NNN,j} - e_{\rm nuc,target,j}|,
\label{eq:DeltaEpsNuc}
\end{equation}
and neutrino loss term 
\begin{equation}
\Delta \epsilon_{\rm \nu} = \frac{1}{N_{\rm test}} \sum_{j=1}^{N_{\rm test}} |\epsilon_{\rm \nu,NNN,j} - \epsilon_{\rm \nu, target,j}|.
\label{eq:DeltaEpsNu}
\end{equation}

\begin{figure*}
\vspace*{-1.5cm}
    \centering
    \subfloat{
        \includegraphics[width=0.485\textwidth]{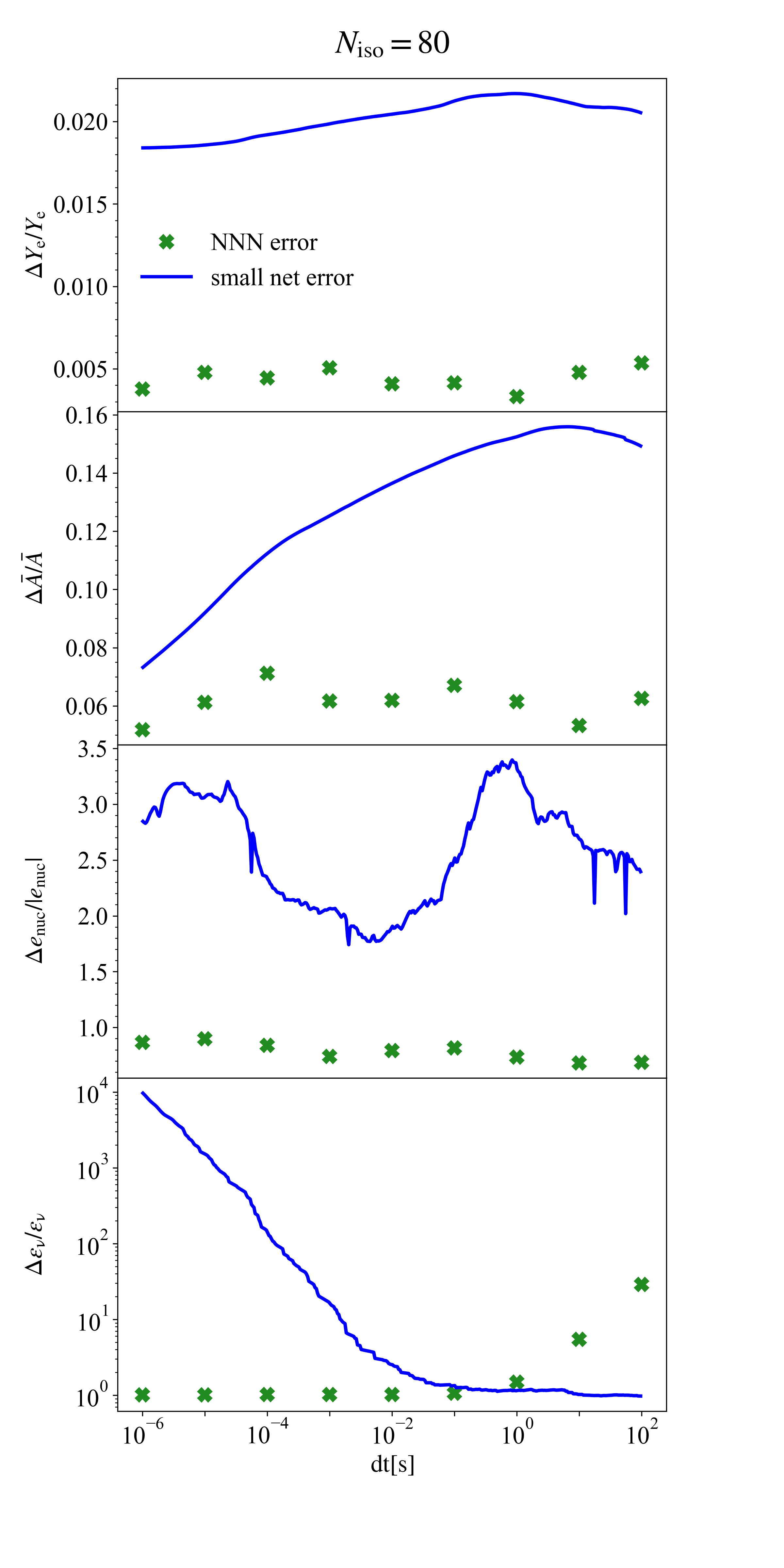}
    }
    \hfill
    \subfloat{
        \includegraphics[width=0.485\textwidth]{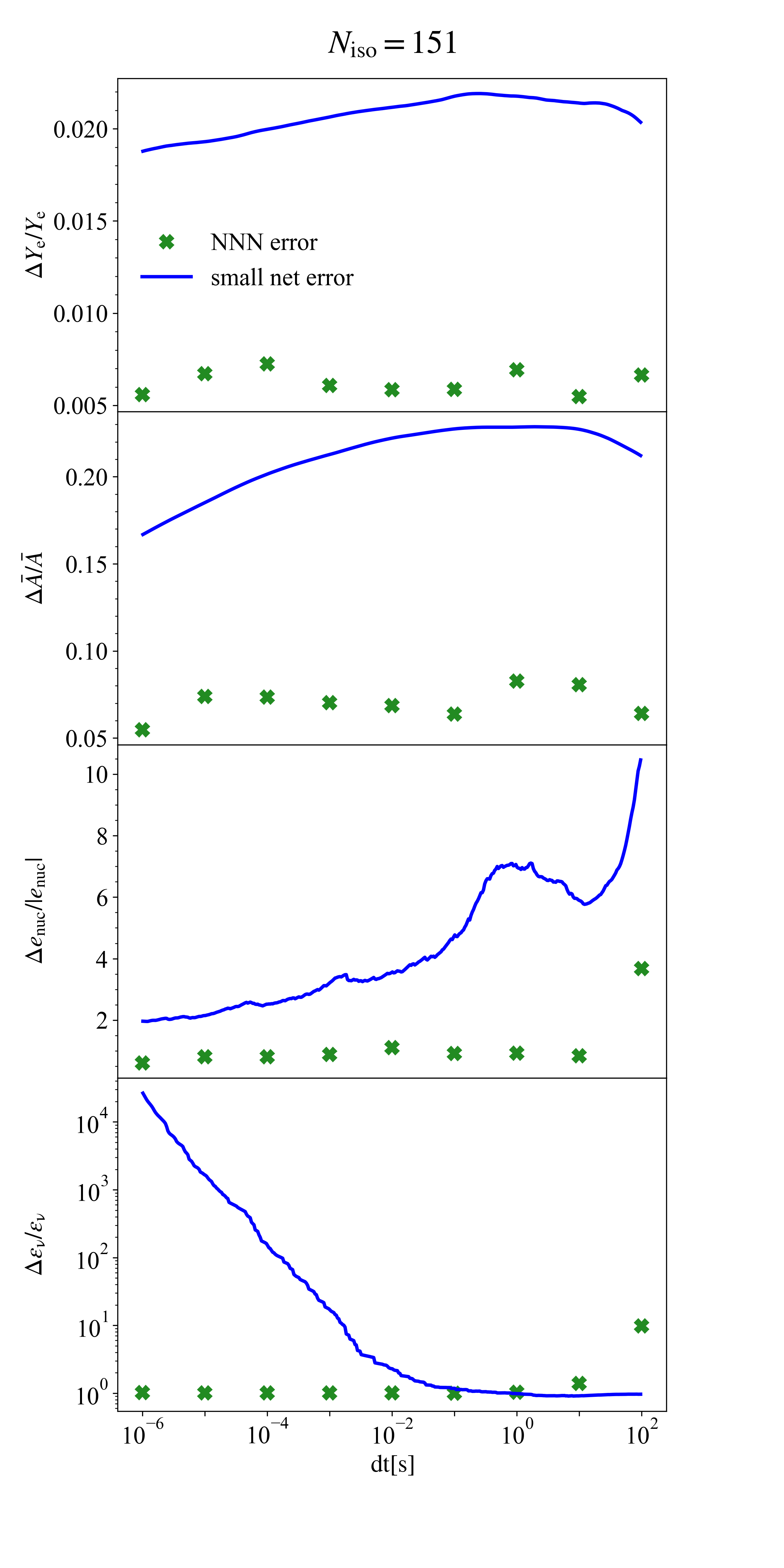}
    }
    \caption{Comparison between the nuclear neural network (NNN) predictions and the results obtained by using a small nuclear reaction network. The NNNs outperform the small net by several factors at small timesteps. The top, upper-middle, lower-middle and bottom panels present the errors in the electron fraction $Y_{\rm e}$,  average mass number $\bar{A}$, nuclear energy generation term $e_{\rm nuc}$ and neutrino-loss term $\epsilon_{\rm \nu}$, respectively. Left panel: the errors for NNNs trained on datasets generated with the nuclear reaction network \textit{mesa\textunderscore 80}. The green X markers are the average errors in the predictions of the NNNs compared to the target values from the $\textsc{bbq}$ runs with \textit{mesa\textunderscore 80} normalized by the average parameter value from \textsc{bbq}. The blue line represents the relative errors that occur while running \textsc{bbq} with the 22-isotope nuclear reaction network \textit{approx21\textunderscore cr60\textunderscore plus\textunderscore co56}. Right panel: same as the left panel for NNNs trained on datasets generated with the nuclear reaction network \textit{mesa\textunderscore 151}.}
    \label{fig:LossesDifferentDt}
\end{figure*}

\begin{figure}
\centering
\vspace*{-0.2cm}
\hspace*{-0.6cm}
\includegraphics[width=0.57\textwidth, trim={0 40 0 0}, clip]{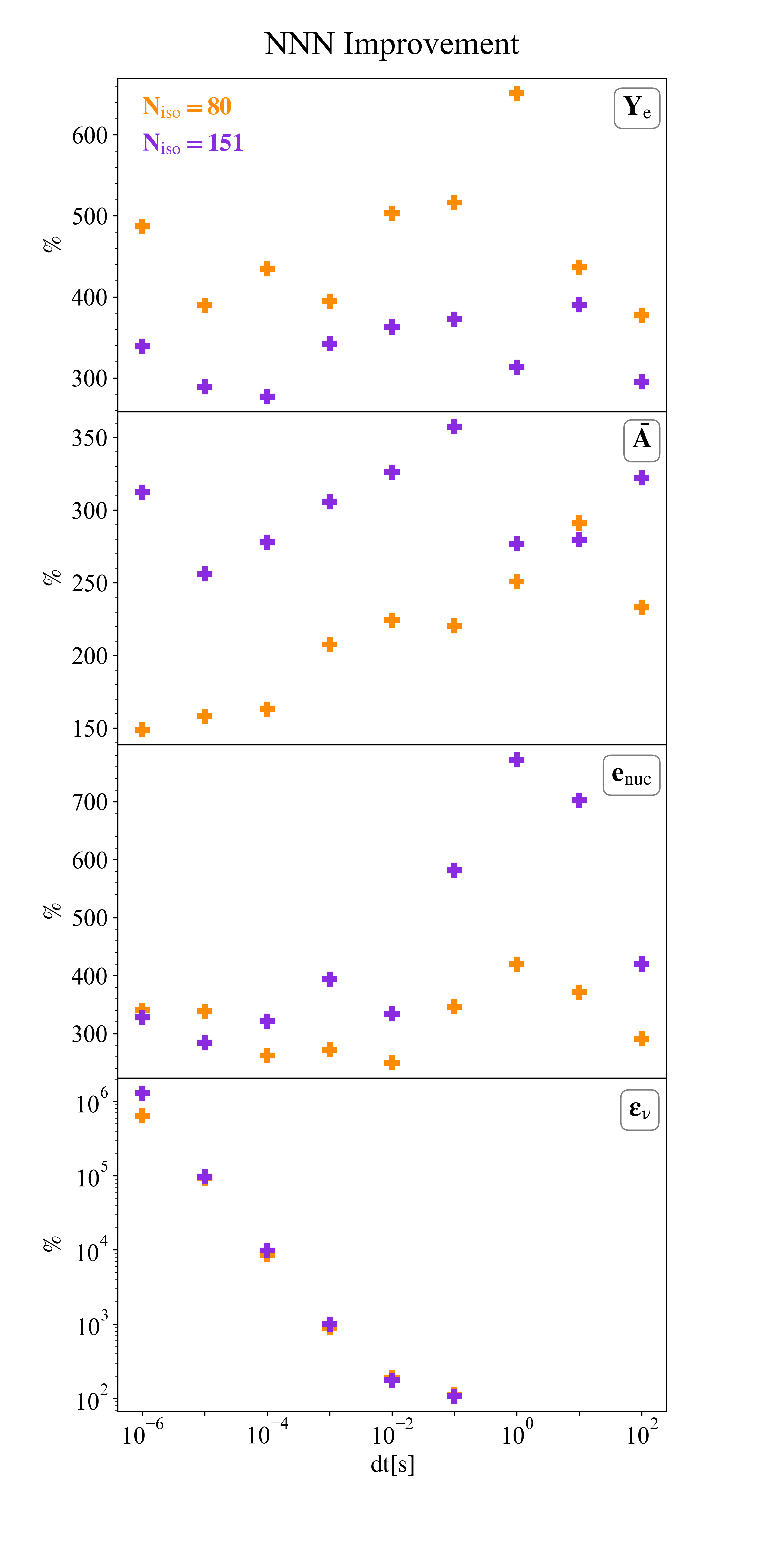}
\captionsetup{skip=0pt}
\caption{Improvement in the results using Nuclear Neural Networks (NNNs) compared to the small nuclear reaction network, obtained by dividing the small net error by the NNN error and converting the result to a percentage. NNN predictions are hundreds of percents more accurate than the small net for most of our explored timesteps. The top,upper-middle, lower-middle and bottom panels present the percentage of improvement in the values of the electron fraction $Y_{\rm e}$, average mass number $\bar{A}$ , nuclear energy generation term $e_{\rm nuc}$ and neutrino-loss term $\epsilon_{\rm \nu}$, respectively.  Orange (purple) pluses are for NNNs trained on \textit{mesa\textunderscore 80} (\textit{mesa\textunderscore 151}) datasets.}
\label{fig:NNN_improvement}
\end{figure}

Fig.~\ref{fig:LossesDifferentDt} shows the relative average errors in the electron fraction $Y_{\rm e}$ (top panels), average mass number (upper-middle panels) (which represents both nuclear numbers, as we will explain later), nuclear energy generation term $e_{\rm nuc}$ (lower-middle panels) and neutrino-loss term $\epsilon_{\rm \nu}$ (bottom panels). The green X markers are the errors in NNN predictions compared to the target values from \textsc{bbq} runs with large nuclear reaction networks computed by equations \ref{eq:DeltaYe} - \ref{eq:DeltaEpsNu}, averaged over 700 tests, and normalized by the average parameter value from \textsc{bbq}, hereafter referred to as `NNN errors'. The blue line marks the normalized relative errors that result from using the nuclear reaction network \textit{approx21\textunderscore cr60\textunderscore plus\textunderscore co56}, calculated by replacing the NNN prediction term in these equations with the result of this small nuclear reaction network, hereafter will be referred to as `small net errors'. The left panel is for the 12-layer NNNs trained on \textit{mesa\textunderscore 80} datasets, and the right panel presents the same results for the 9-layer NNNs trained on \textit{mesa\textunderscore 151} datasets. 

As shown in the top panels of Fig.~\ref{fig:LossesDifferentDt}, for both nuclear reaction networks the NNN errors in the electron fraction $Y_{\rm e}$ are significantly smaller than the small net errors, remaining below $0.75\%$ across all the timesteps we explored. The time evolution of $Y_{\rm e}$ leads to a growth in the small net errors at short timescales as \textit{approx21\textunderscore cr60\textunderscore plus\textunderscore co56} handles weak interactions through the single compound effective reaction ${^{56}\mathrm{Fe} + 4n + 2e^{-} \rightarrow ^{60}\mathrm{Cr}+2\nu_{\rm e}}$, which lowers the electron fraction before the electron captures and $\beta^{+}$ decays implemented in larger networks become important. Overall, the small net errors in $Y_{\rm e}$ are in accordance with the previous studies of \citealt{Farmeretal2016} and \citealt{Renzoetal2024} that examined full stellar models. The NNN errors in $Y_{\rm e}$, however, depend solely on how well the NNN manages to reproduce the composition of the large nuclear reaction networks, and remain constant on average over time. We find that the NNNs outperform the small nuclear reaction network by $280-400\%$ in the case of \textit{mesa\textunderscore 151} (top panel of Fig.~\ref{fig:NNN_improvement}; purple pluses) and by $390-660\%$ for \textit{mesa\textunderscore 80} (top panel of Fig.~\ref{fig:NNN_improvement}; orange pluses).

The average mass number (upper-middle panels of Fig.~\ref{fig:LossesDifferentDt}), which also depends directly on the composition shows a similar behavior. The NNNs achieve improvements of $150-290\%$ and $260-360\%$ over the small net relative to the values computed using \textit{mesa\textunderscore 80} (orange pluses in the upper-middle panel of Fig.~\ref{fig:NNN_improvement}) and \textit{mesa\textunderscore 151} (purple  pluses in the upper-middle panel of Fig.~\ref{fig:NNN_improvement}), respectively. Performing the same analysis on the average atomic number, we find indistinguishable errors due to the close relation between these two quantities ($\bar{A}\simeq 2\bar{Z}$).

For the energy terms, the small net errors are large, since it does not account for weak nuclear reactions at advanced burning stages. In the case of the nuclear energy generation term $e_{\rm nuc}$ (lower-middle panel of Fig.~\ref{fig:LossesDifferentDt}), the small net errors are very high throughout silicon core burning, reaching $200-1000 \%$ error. While the NNN errors are large as well, ranging approximately around $70-90\%$ error in the case of  \textit{mesa\textunderscore 80} (left panel) and mostly around $60-100\%$ (except for our latest timestep $dt=100\s$ where the NNN error is $360\%$) for \textit{mesa\textunderscore 151} (right panel), they perform better than the smaller network, maintaining an error at least twice as small for all our explored timesteps. For \textit{mesa\textunderscore 80} the NNN achieves improvements of $250-450\%$, with an even higher performance increase of $280-750\%$ in the case of \textit{mesa\textunderscore 151} (lower-middle panel of Fig.~\ref{fig:NNN_improvement}).

In the case of the neutrino-loss term (bottom panel of Fig.~\ref{fig:LossesDifferentDt}), the behavior of the error is similar for both nuclear reaction networks. At short timesteps the NNN errors (around $100\%$ for both \textit{mesa\textunderscore 80} and \textit{mesa\textunderscore 151}) are negligible compared to the small net errors, with NNNs outperforming the small net by $100\%$ to $10^{6}\%$ (bottom panel of Fig.~\ref{fig:NNN_improvement}),but both relative errors are large. They become comparable around $dt=0.1 \s$. After that, the small net error plateaus and the NNN error increases, resulting in larger NNN errors compared to small net error at long times. We note that since the NNNs emulate the results of \textsc{bbq} and have no information on the underlying nuclear physics of the system, the sum of the nuclear and neutrino energy densities predicted by the NNNs will not amount to the sum over the rest mass difference between reactants and products for all nuclear reactions. Therefore, it is necessary to train the NNNs to predict both $e_{\rm nuc}$ and $\epsilon_{\rm \nu}$ for given initial conditions, rather than predicting one and inferring the other from equation \ref{eq:EpsNucEpsNu}. 

\begin{figure*}
\begin{center}
\includegraphics[width=1\textwidth]{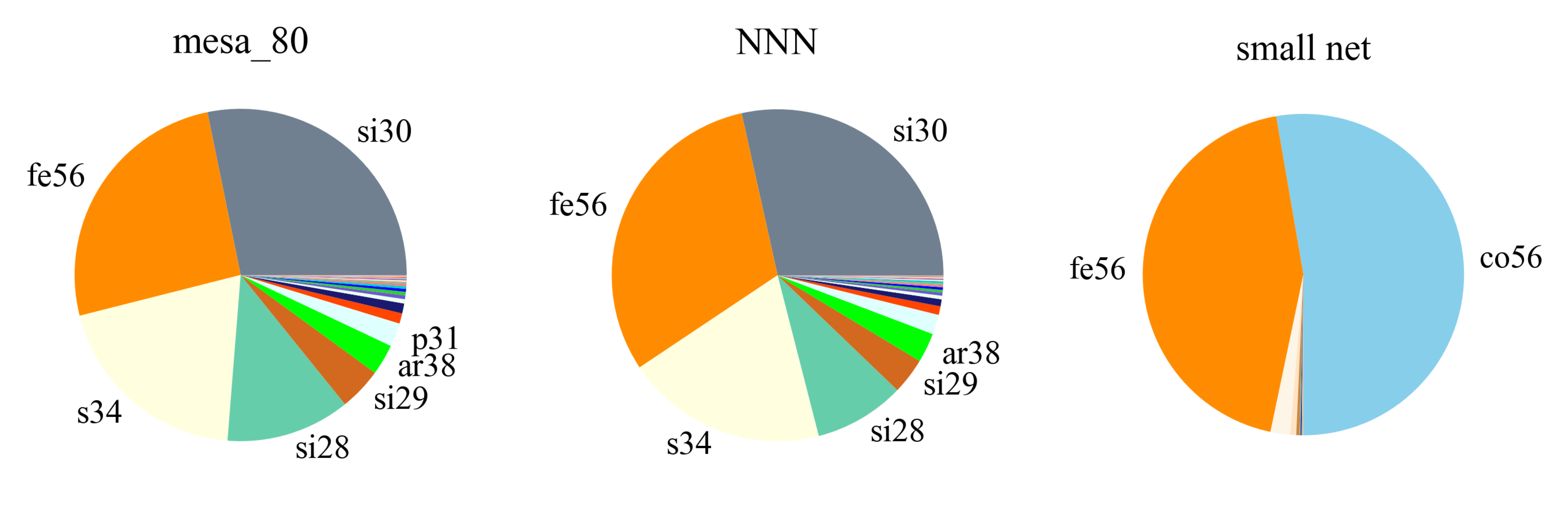}
\caption{Comparison between the performance of our nuclear neural network (NNN) in predicting a stellar composition and the isotopic abundances obtained using a small nuclear reaction network. We show the compositions of a stellar burning region evolved at a constant temperature of $T=5\times 10^{9} \K$ and a constant density of $\rho =2\times 10^{8} \K$ during $dt=0.1 \s$. Left panel: the composition resulting from a \textsc{bbq} run performed with the large nuclear reaction network \textit{mesa\textunderscore 80} (the target composition). Middle panel: the composition predicted using our trained NNN model for $dt=0.1 \s$. Right panel: the composition resulting from a \textsc{bbq} run performed with the small nuclear reaction network \textit{approx21\textunderscore cr60\textunderscore plus\textunderscore co56}.   }
\label{fig:PieCharts}
\end{center}
\end{figure*}

Examining the NNN predictions of the compositions, we find that while in the best cases they yield the exact same abundances as obtained with the large nuclear reaction networks for most of the isotopes, the worst predictions give compositions completely unrelated to the target values. In Fig.~\ref{fig:PieCharts} we compare between the NNN predictions of a 'fiducial' stellar composition with an electron fraction error $\Delta Y_{\rm e}/Y_{\rm e}=0.004$ that corresponds to the average value found in the top panel of Fig.~\ref{fig:LossesDifferentDt} (middle panel) \footnote{We selected the 'fiducial' composition by first filtering those with an average error in $Y_{\rm e}$, and then choosing the composition whose errors in the remaining parameters were closest to the respective average values, though some deviations exist. }, and the results obtained by running the small net (right panel). We can see that while in both cases the isotope abundances deviate significantly from the composition obtained using \textit{mesa\textunderscore 80} (left panel), the NNN performs notably better in predicting the abundances of the main isotopes. 

\section{Applicability to Stellar Evolution Modeling}
\label{sec:applicability}

\begin{figure}
\begin{center}
\hspace*{-0.5cm}
\includegraphics[width=0.48\textwidth]{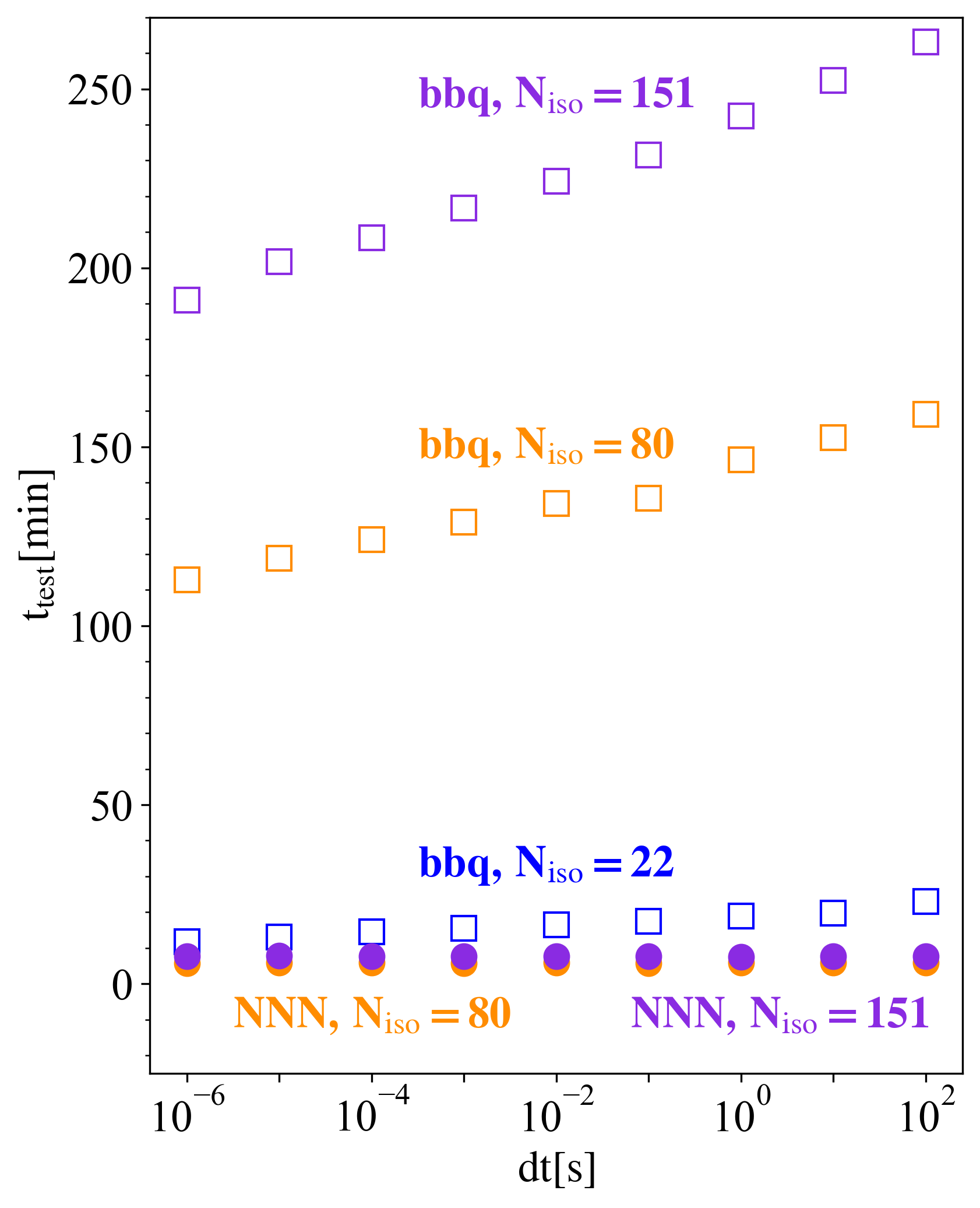}
\caption{Comparison between the computation time required to determine the composition and energy terms using nuclear neural networks (NNNs) and the accurate ordinary differential equations (ODEs) solvers. Empty orange (purple) squares represent the time it takes \textsc{bbq} solving the nucleosynthesis ODEs for all test datasets ($\simeq700$ output files) with a nuclear reaction network that consists of 80 (151) isotopes. Filled orange (purple) circles indicate the time needed to predict these quantities using NNN models trained on 80 (151) isotopes. For comparison, we plot the computation time required for solving the ODEs using the small net (22 isotopes).  }
\label{fig:test_time_vs_dt}
\end{center}
\end{figure}

Building on this proof-of-concept, the next step would be to integrate the NNNs into a stellar evolution code and assess their performance in full stellar evolution simulations. Replacing accurate ODE solvers with NNN trained models has the potential to significantly accelerate modeling of massive stars. Fig.~\ref{fig:test_time_vs_dt} compares the time required to produce the testing datasets with \textsc{bbq} using a nuclear reaction networks with 80 (empty orange squares) and 151 (empty purple squares) isotopes, to the time it takes the NNN trained models to predict the compositions and energy terms of these datasets (filled orange circles for NNNs trained on \textit{mesa\textunderscore 80} datasets and filled purple circles for NNNs trained on \textit{mesa\textunderscore 151} datasets) for the range of timesteps we explored in this study. The addition of more isotopes (and consequently more ODEs and terms in each equation; see equation \ref{eq:NucODEs}) significantly increases the time required by \textsc{bbq} to generate the test datasets, whereas the NNN prediction time is less affected. Furthermore, while the running time of \textsc{bbq} increases with larger timesteps due to more prominent compositional changes that complicate the ODE solving, NNN predictions are performed in nearly constant computational time, regardless of the timestep size. 

Overall, we can see that in the case of \textit{mesa\textunderscore 80} NNN is more efficient than \textsc{bbq} in finding the composition and energy terms by a factor of 20-27, and for \textit{mesa\textunderscore 151} the efficiency grows by 25-34 in favor of NNNs (without accounting for the one-time computational cost of generating the training sets). Comparing to the time required to produce the same \textsc{bbq} test sets with the small, 22-isotopes nuclear reaction network (blue empty squares), we find NNNs are still 1.5-4 times faster. Since in late stages of modeling massive stars nuclear burning takes the vast majority of computational time, implementing NNNs in stellar evolution codes could significantly reduce the computation time needed to evolve models with large nuclear reaction networks. More importantly, since NNNs emulate the nucleosynthesis datasets rather than solving stiff ODEs, they could improve the numerical stability of stellar models and lower their likelihood to crash by reducing the need for extremely short timesteps.

This approach could facilitate the development of improved CCSN progenitor models, allowing for a more comprehensive exploration of the SN explosion mechanism. In particular, this will be of crucial importance for multidimensional hydrodynamical simulations of SN explosions, which typically rely on stellar evolution models with small nuclear reaction networks as their starting point (e.g., \citealt{Fields2022}; \citealt{Fujibayashietal2024}; \citealt{WangPan2024}; \citealt{Zingaleetal2024}). Furthermore, the models trained on small timesteps could be useful in emulating nucleosynthesis during the explosion itself, though might require larger nuclear reaction networks.  While covering the parameter space regime of the most common CCSN and black hole progenitors (see Fig.~\ref{fig:ParamasSelection}), we note that lower mass stars that explode in electron capture SNe \footnote{We note that modeling electron capture SNe requires large nuclear reaction networks with an extensive net of weak reactions (see e.g., \citealt{Jonesetal2013}).} have lower central temperatures throughout the final stages of their evolution (e.g., \citealt{Wangetal2024}), and higher mass stars that undergo pulsational pair instability SNe typically exhibit lower central densities (e.g., \citealt{Woosley2017}; \citealt{Renzoetal2020}).  As a result, our trained models are unsuitable for these scenarios; our work, however, could be readily extended to incorporate them by using the same method (section~\ref{sec:method}). 

Integrating NNN trained models into stellar evolution codes would require to split the burning from the stellar structure and mixing evolution, as already possible in \textsc{mesa} for high temperatures and densities using the split-operator \citep{Jermynetal2023}. The NNNs would then receive the isotopic abundances of each burning region at a given timestep after rotation and convection mixed the composition obtained by NNNs in the previous timestep across regions. The burner would need to fix the timestep and select the NNN-trained model most compatible with that duration, which will be the model trained on the largest timestep constrained by nuclear physics. If the chosen timestep is not an exact multiple of any training timestep, the burner will interpolate between the two nearest times to obtain the nucleosynthesis results. The NNNs would be used only for burning regions with parameters within the training regime, as fully-connected neural networks typically perform poorly when extrapolating (e.g., \citealt{Kimetal2020}). It is necessary to train models on additional timesteps and nuclear physics input to cover a broader parameter space.

To incorporate NNNs as surrogate models for nuclear reaction networks in stellar evolution codes it is crucial to conduct parallel tests alongside accurate solvers and assess the accumulation of NNN errors over multiple timesteps. When evaluating the performance of the NNNs over a sequence of 1000 timesteps at constant temperature and density, we find that the errors in $Y_{\rm e}$ generally remain smaller than those of the small network, while the errors in the other quantities presented in Fig.~\ref{fig:LossesDifferentDt} become comparable and can exceed those of the small net. We note, however, that in practice the NNNs would always be called using the largest possible trained timestep, implying they will not be applied recursively to their own outputs. Between calls, the temperature and density evolve, and the composition is mixed across burning regions. Analyzing the distribution of errors across the parameter space and their temporal evolution within a stellar evolution code and performing a comparison with errors from conventional ODE solvers would provide valuable insight into the regimes where the NNNs underperform and help in improving the models. Apart from monitoring the compounding errors in the predictions presented in Fig. \ref{fig:LossesDifferentDt}, it is essential to evaluate how well energy is conserved in the stellar model, as NNNs emulate the energy terms without explicitly enforcing thermodynamic consistency. 

Generating more targeted training sets by obtaining abundances from full stellar evolution models on a large scale could improve the accuracy of the NNN predictions and bring us closer to their implementation. Extracting compositions from cores of 150 \textsc{mesa} models of masses $15-30 M\odot$, each evolved using the small nuclear reaction network, as initial conditions for \textsc{bbq} resulted in NNNs with errors several times larger than those of the small net in predicting the nuclear energy generation term. A possible solution could be modeling a smaller number of reliable CCSN progenitors and varying their compositions around the main isotopes to obtain big enough training sets. However, this would still require stellar tracks from a large amount of stellar models throughout silicon core burning, which is computationally challenging due to the stiffness of the nucleosynthesis ODEs. A special emphasis should be given to the boundary regimes between NNNs and the accurate solvers to ensure a continuous smooth transition. This, along building the infrastructure to replace ODE solver with NNN trained models in \textsc{mesa}, is a subject of future studies.

Other important uncertainties concerning nucleosynthesis in massive stars are currently overlooked by stellar evolution codes. The overall uncertainty in the measurement of reaction rates (see \citealt{Fanetal2022} and \citealt{Smithetal2023} for frameworks designed to explore nuclear reaction networks uncertainty) would have a great influence on stellar evolution outcomes (e.g., \citealt{Fieldsetal2018}). Variations in the reaction rate of carbon burning, for instance, can result in a significant impact on the nucleosynthesis of heavier elements and the explodability of massive stars (e.g., \citealt{Farmeretal2019}; \citealt{Costaetal2021}; \citealt{Xinetal2023}; \citealt{Dumontetal2024} \citealt{Xinetal2025}). Another example is the contribution of excited states in out of equilibrium nuclei to nuclear energy generation (e.g., \citealt{MischMumpower2024}) and neutrino energies (e.g., Farag et al.2025 in prep) during advanced burning stages. The difficulty in accounting for some of these effects presents a significant challenge to accurately simulating stellar nucleosynthesis in CCSN progenitors. A systematic study of the uncertainties to compare their effects to the small net errors is beyond our present scope. 

\section{Summary}
\label{sec:Summary}

In this study we developed neural network-based emulators (Section~\ref{sec:method}) and conducted a proof-of-concept to evaluate their ability to predict the composition and energy generation during late burning stages of massive stars (Section~\ref{sec:results}) to lay the ground work for future integration into stellar evolution codes (Section~\ref{sec:applicability}). The large number of isotopes required to accurately model nuclear burning presents a significant numerical challenge, limiting our ability to study these evolutionary stages and to produce reliable progenitor models for CCSN simulations. To overcome this, we designed neural networks and trained them on datasets produced using the nuclear module of the stellar evolution code \textsc{mesa}, accessed through the wrapper \textsc{bbq}. 

We defined the boundaries of our training sets to be in the temperature regime ${10^{9.2} \K <T<10^{9.9} \K}$ and density regime ${10^{7} \g \cm ^{-3}<\rho < 10^{9} \g \cm ^{-3}}$ by examining several \textsc{mesa} models computed with large nuclear reaction networks, and used the same simulations to determine the timesteps for training our NNNs, ensuring their relevance to stellar evolution modeling (Fig.~\ref{fig:ParamasSelection}). We used \textsc{PyTorch lightning} to design NNNs that receive the temperature (on a logarithmic scale), density (on a logarithmic scale) and composition (on a linear scale) of a burning region as input, map them through multiple hidden layers, and output the updated composition, normalized nuclear energy generation term, and normalized neutrino-loss term after a given timestep (Fig.~\ref{fig:NNNarchitecture}). 

We tested the predictions of the NNNs for the composition against the target values from \textsc{bbq} runs and presented the results in Fig.~\ref{fig:AbundanceLossDensityMap}. Due to the large errors, we caution against using them in studies where the details of the composition are important (see also Fig.~\ref{fig:PieCharts}). For the electron fraction computed using this predicted composition, however, the NNNs achieve an average percent difference of $0.4-0.75\%$ (top panels of Fig.~\ref{fig:LossesDifferentDt}), which is $280-660\%$ higher than the accuracy obtained with the commonly used small nuclear reaction network (top panel of Fig.~\ref{fig:NNN_improvement}). As the electron fraction is crucial in determining the explodability of massive stars, an improvement in several factors could result in a different outcome that better reflects reality. 

Examining the average nuclear numbers and the nuclear energy generation rate for datasets produced with 80 and 151 isotopes, we found that across all explored timesteps the NNNs produced errors  at least 1.5 times smaller than those from the smaller network ( upper middle and lower-middle panels of Figs.~\ref{fig:LossesDifferentDt} and \ref{fig:NNN_improvement}, respectively). For the rate of energy lost by neutrinos from weak nuclear reactions, the bottom panels of Fig.~\ref{fig:LossesDifferentDt} and \ref{fig:NNN_improvement} show that the NNNs perform $100-10^{6}\%$ better than the small net for short timesteps, while resulting in comparable and larger errors for $dt \gtrsim 0.1 \s$. 

Comparing between the typical times for obtaining the compositions and energy generation terms by solving the nucleosynthesis ODEs with \textsc{bbq} and predicting them with our NNN trained models, we found that NNNs have the potential of accelerating stellar evolution simulations performed with large nuclear reaction networks by a factor of 20-34 (Fig.~\ref{fig:test_time_vs_dt}), making them a promising tool for generating more reliable CCSN progenitors on larger scales. Before integrating them into stellar evolution simulations, however, it is important to assess the accumulation of NNN errors throughout silicon core burning, and to compare the improvement in their results with respect to the commonly used small nuclear reaction networks and other nuclear physics uncertainties.

\section*{Acknowledgments}

We thank Alexander Heger, Dmitry Shishkin, Kaze Wong, Noam Soker, Sultan Hassan and Vladimir Kalnitsky for helpful discussion. We thank an anonymous referee for useful suggestions and insights that greatly enhanced the quality of this manuscript. We thank the Kavli Foundation and the Max Planck Institute for Astrophysics for supporting the 2023 Kavli Summer Program during which much of this work was completed. We thank the Center for Computational Astrophysics for granting us computing time. AG acknowledges support from the Miriam and Aaron Gutwirth Fellowship, the Steward Observatory Fellowship in Theoretical and Computational Astrophysics, the IAU-Gruber Fellowship, and the CHE Fellowship.

\section*{Data Availability}
\label{sec:DataAvailability}
The training sets, trained NNN models, test datasets, \textsc{mesa} stellar models, and \textsc{python} scripts used for the analysis performed in this manuscript are publicly available on \href{https://zenodo.org/records/14873443?token=eyJhbGciOiJIUzUxMiJ9.eyJpZCI6IjJiMmM3ZDgyLWNiYjctNDA0MS04Y2U3LTBlMjJmOWQ5ZTY0ZCIsImRhdGEiOnt9LCJyYW5kb20iOiIzNDFmMTgzODE4MDMzZTU0NTU4YzJiOGI1YmY2MzgzYyJ9.rHV9FJScD_-KIcdJN1NRgTTPhsaGxTlNs3QEu4m5TzV4rOZltiKxNxf5qY_cXxjpZs8OpQVvVNRNh0xO6ViWoA}{10.5281/zenodo.14873443}, and detailed documentation is available in this \href{https://sites.google.com/view/aldanagrichener/nuclear-neural-networks?authuser=0}{guide}. The complete \textsc{bbq} runs that include additional timesteps for training the NNNs and datasets for extended density ranges, comprising a total of 4TB, will be shared upon request from the corresponding author.   

\textit{Software}: \textsc{bbq} \citep{bbqFromGitHub}, \textsc{matplotlib} \citep{Hunteretal2007}, \textsc{mesa-23.05.1} (\citealt{Paxtonetal2011}; \citealt{Paxtonetal2013}; \citealt{Paxtonetal2015}; \citealt{Paxtonetal2018}; \citealt{Paxtonetal2019}; \citealt{Jermynetal2023}), \textsc{numpy} \citep{vanderWalt2011}, \textsc{pandas} \citep{Pandas2024}, \textsc{python} \citep{python2009}, \textsc{pytorch Lightning} \citep{FalconBorovec2020}, \textsc{scipy} \citep{VirtanenGommers2020}. 


\bibliography{refs} 

\appendix

\setcounter{figure}{0}
\renewcommand{\thefigure}{A\arabic{figure}}

\section{Isotope List of the Nuclear Reaction Networks}
\label{sec:AppendixIsos}

\begin{longtable}{|c|c|c|}
\hline
\textbf{\textit{approx21\textunderscore cr60\textunderscore plus\textunderscore co56}} & \textbf{\textit{mesa\_80}} & \textbf{\textit{mesa\_151}} \\ \hline
\endfirsthead
\hline
\textbf{\textit{approx21\textunderscore cr60\textunderscore plus\textunderscore co56}} & \textbf{\textit{mesa\_80}} & \textbf{\textit{mesa\_151}} \\ \hline
\endhead
n & n & n \\ \hline
$^{1}\mathrm{H}$ & $^{1}\mathrm{H}$, $^{2}\mathrm{H}$ & $^{1}\mathrm{H}$, $^{2}\mathrm{H}$ \\ \hline
p & & \\ \hline
$^{3}\mathrm{He}$ & $^{3}\mathrm{He}$ & $^{3}\mathrm{He}$ \\ \hline
$^{4}\mathrm{He}$ & $^{4}\mathrm{He}$ & $^{4}\mathrm{He}$ \\ \hline
 & $^{7}\mathrm{Li}$ & $^{6}\mathrm{Li}$, $^{7}\mathrm{Li}$ \\ \hline
 & $^{7}\mathrm{Be}$, $^{9}\mathrm{Be}$, $^{10}\mathrm{Be}$ & $^{7}\mathrm{Be}$, $^{9}\mathrm{Be}$, $^{10}\mathrm{Be}$ \\ \hline
 & $^{8}\mathrm{B}$ & $^{8}\mathrm{B}$, $^{10}\mathrm{B}$, $^{11}\mathrm{B}$ \\ \hline
$^{12}\mathrm{C}$ & $^{12}\mathrm{C}$, $^{13}\mathrm{C}$ & $^{12}\mathrm{C}$, $^{13}\mathrm{C}$ \\ \hline
$^{14}\mathrm{N}$ & $^{13}\mathrm{N}$, $^{14}\mathrm{N}$, $^{15}\mathrm{N}$ & $^{13}\mathrm{N}$, $^{14}\mathrm{N}$, $^{15}\mathrm{N}$, $^{16}\mathrm{N}$ \\ \hline
$^{16}\mathrm{O}$ & $^{14}\mathrm{O}$, $^{15}\mathrm{O}$, $^{16}\mathrm{O}$, $^{17}\mathrm{O}$, $^{18}\mathrm{O}$ & $^{15}\mathrm{O}$, $^{16}\mathrm{O}$, $^{17}\mathrm{O}$, $^{18}\mathrm{O}$, $^{19}\mathrm{O}$ \\ \hline
 & $^{17}\mathrm{F}$, $^{18}\mathrm{F}$, $^{19}\mathrm{F}$ & $^{17}\mathrm{F}$, $^{18}\mathrm{F}$, $^{19}\mathrm{F}$, $^{20}\mathrm{F}$ \\ \hline
$^{20}\mathrm{Ne}$ & $^{18}\mathrm{Ne}$, $^{19}\mathrm{Ne}$, $^{20}\mathrm{Ne}$, $^{21}\mathrm{Ne}$, $^{22}\mathrm{Ne}$ & $^{19}\mathrm{Ne}$, $^{20}\mathrm{Ne}$, $^{21}\mathrm{Ne}$, $^{22}\mathrm{Ne}$, $^{23}\mathrm{Ne}$ \\ \hline
 & $^{21}\mathrm{Na}$, $^{22}\mathrm{Na}$, $^{23}\mathrm{Na}$, $^{24}\mathrm{Na}$ & $^{21}\mathrm{Na}$, $^{22}\mathrm{Na}$, $^{23}\mathrm{Na}$, $^{24}\mathrm{Na}$ \\ \hline
$^{24}\mathrm{Mg}$ & $^{23}\mathrm{Mg}$, $^{24}\mathrm{Mg}$, $^{25}\mathrm{Mg}$, $^{26}\mathrm{Mg}$ & $^{23}\mathrm{Mg}$, $^{24}\mathrm{Mg}$, $^{25}\mathrm{Mg}$, $^{26}\mathrm{Mg}$, $^{27}\mathrm{Mg}$ \\ \hline
 & $^{25}\mathrm{Al}$, $^{26}\mathrm{Al}$, $^{27}\mathrm{Al}$ & $^{25}\mathrm{Al}$, $^{26}\mathrm{Al}$, $^{27}\mathrm{Al}$, $^{28}\mathrm{Al}$ \\ \hline
$^{28}\mathrm{Si}$ & $^{27}\mathrm{Si}$, $^{28}\mathrm{Si}$, $^{29}\mathrm{Si}$, $^{30}\mathrm{Si}$ & $^{27}\mathrm{Si}$, $^{28}\mathrm{Si}$, $^{29}\mathrm{Si}$, $^{30}\mathrm{Si}$, $^{31}\mathrm{Si}$, $^{32}\mathrm{Si}$, $^{33}\mathrm{Si}$ \\ \hline
 & $^{30}\mathrm{P}$, $^{31}\mathrm{P}$ & $^{30}\mathrm{P}$, $^{31}\mathrm{P}$, $^{32}\mathrm{P}$, $^{33}\mathrm{P}$, $^{34}\mathrm{P}$ \\ \hline
$^{32}\mathrm{S}$ & $^{31}\mathrm{S}$, $^{32}\mathrm{S}$, $^{33}\mathrm{S}$, $^{34}\mathrm{S}$ & $^{31}\mathrm{S}$, $^{32}\mathrm{S}$, $^{33}\mathrm{S}$, $^{34}\mathrm{S}$, $^{35}\mathrm{S}$, $^{36}\mathrm{S}$, $^{37}\mathrm{S}$ \\ \hline
 & $^{35}\mathrm{Cl}$ & $^{35}\mathrm{Cl}$, $^{36}\mathrm{Cl}$, $^{37}\mathrm{Cl}$, $^{38}\mathrm{Cl}$ \\ \hline
$^{36}\mathrm{Ar}$ & $^{35}\mathrm{Ar}$, $^{36}\mathrm{Ar}$, $^{37}\mathrm{Ar}$, $^{38}\mathrm{Ar}$ & $^{36}\mathrm{Ar}$, $^{37}\mathrm{Ar}$, $^{38}\mathrm{Ar}$, $^{39}\mathrm{Ar}$, $^{40}\mathrm{Ar}$, $^{41}\mathrm{Ar}$ \\ \hline
 & $^{39}\mathrm{K}$ & $^{39}\mathrm{K}$, $^{40}\mathrm{K}$, $^{41}\mathrm{K}$, $^{42}\mathrm{K}$ \\ \hline
$^{40}\mathrm{Ca}$ & $^{39}\mathrm{Ca}$, $^{40}\mathrm{Ca}$, $^{42}\mathrm{Ca}$ & $^{40}\mathrm{Ca}$, $^{42}\mathrm{Ca}$, $^{43}\mathrm{Ca}$, $^{44}\mathrm{Ca}$, $^{45}\mathrm{Ca}$, $^{46}\mathrm{Ca}$, $^{47}\mathrm{Ca}$, $^{48}\mathrm{Ca}$, $^{49}\mathrm{Ca}$ \\ \hline
 & $^{43}\mathrm{Sc}$ & $^{43}\mathrm{Sc}$, $^{44}\mathrm{Sc}$, $^{45}\mathrm{Sc}$, $^{46}\mathrm{Sc}$, $^{47}\mathrm{Sc}$, $^{48}\mathrm{Sc}$, $^{49}\mathrm{Sc}$ \\ \hline
$^{44}\mathrm{Ti}$ & $^{44}\mathrm{Ti}$, $^{45}\mathrm{Ti}$, $^{46}\mathrm{Ti}$ & $^{44}\mathrm{Ti}$, $^{45}\mathrm{Ti}$, $^{46}\mathrm{Ti}$, $^{47}\mathrm{Ti}$, $^{48}\mathrm{Ti}$, $^{49}\mathrm{Ti}$, $^{50}\mathrm{Ti}$, $^{51}\mathrm{Ti}$ \\ \hline
 & $^{47}\mathrm{V}$ & $^{47}\mathrm{V}$, $^{48}\mathrm{V}$, $^{49}\mathrm{V}$, $^{50}\mathrm{V}$, $^{51}\mathrm{V}$, $^{52}\mathrm{V}$ \\ \hline
$^{48}\mathrm{Cr}$ & $^{48}\mathrm{Cr}$, $^{49}\mathrm{Cr}$, $^{50}\mathrm{Cr}$ & $^{48}\mathrm{Cr}$, $^{49}\mathrm{Cr}$, $^{50}\mathrm{Cr}$, $^{51}\mathrm{Cr}$, $^{52}\mathrm{Cr}$, $^{53}\mathrm{Cr}$, $^{54}\mathrm{Cr}$, $^{55}\mathrm{Cr}$ \\ \hline
 & $^{51}\mathrm{Mn}$ & $^{51}\mathrm{Mn}$, $^{52}\mathrm{Mn}$, $^{53}\mathrm{Mn}$, $^{54}\mathrm{Mn}$, $^{55}\mathrm{Mn}$, $^{56}\mathrm{Mn}$ \\ \hline
$^{52}\mathrm{Fe}$, $^{54}\mathrm{Fe}$, $^{56}\mathrm{Fe}$ & $^{52}\mathrm{Fe}$, $^{53}\mathrm{Fe}$, $^{54}\mathrm{Fe}$, $^{56}\mathrm{Fe}$ & $^{52}\mathrm{Fe}$, $^{53}\mathrm{Fe}$, $^{54}\mathrm{Fe}$, $^{55}\mathrm{Fe}$, $^{56}\mathrm{Fe}$, $^{57}\mathrm{Fe}$, $^{58}\mathrm{Fe}$, $^{59}\mathrm{Fe}$, $^{60}\mathrm{Fe}$, $^{61}\mathrm{Fe}$ \\ \hline
$^{56}\mathrm{Co}$ & $^{55}\mathrm{Co}$, $^{56}\mathrm{Co}$ & $^{55}\mathrm{Co}$, $^{56}\mathrm{Co}$, $^{57}\mathrm{Co}$, $^{58}\mathrm{Co}$, $^{59}\mathrm{Co}$, $^{60}\mathrm{Co}$, $^{61}\mathrm{Co}$ \\ \hline
$^{56}\mathrm{Ni}$ & $^{56}\mathrm{Ni}$, $^{57}\mathrm{Ni}$, $^{58}\mathrm{Ni}$, $^{59}\mathrm{Ni}$ & $^{56}\mathrm{Ni}$, $^{57}\mathrm{Ni}$, $^{58}\mathrm{Ni}$, $^{59}\mathrm{Ni}$, $^{60}\mathrm{Ni}$, $^{61}\mathrm{Ni}$, $^{62}\mathrm{Ni}$, $^{63}\mathrm{Ni}$, $^{64}\mathrm{Ni}$, $^{65}\mathrm{Ni}$ \\ \hline
$^{60}\mathrm{Cr}$ & & \\ \hline
 & $^{59}\mathrm{Cu}$ & \\ \hline
 & $^{60}\mathrm{Zn}$ & \\ \hline
\end{longtable}
\vspace{-1pt} 
\begin{center}
   \noindent\textbf{Table 1: List of isotopes included in the nuclear reaction networks used in this study.} 
\end{center}
\label{table:Isos}

Table \ref{table:Isos} presents the set of isotopes composing the nuclear reaction networks \textit{approx21\textunderscore cr60\textunderscore plus\textunderscore co56}, \textit{mesa\_80} and \textit{mesa\_151} (left to right).  \textit{approx21\textunderscore cr60\textunderscore plus\textunderscore co56} is obtained by adding co56 to the commonly used nuclear network \textit{approx21}, and replacing cr56 with cr60. It is designed to obtain central electron fractions that match more closely the results from larger networks

\newpage

\section{A Bug in Some Large Nuclear Reaction Networks in MESA}
\label{sec:AppendixMesaBug}

\begin{figure*}[h]
\begin{center}
\includegraphics[width=1.0\textwidth]{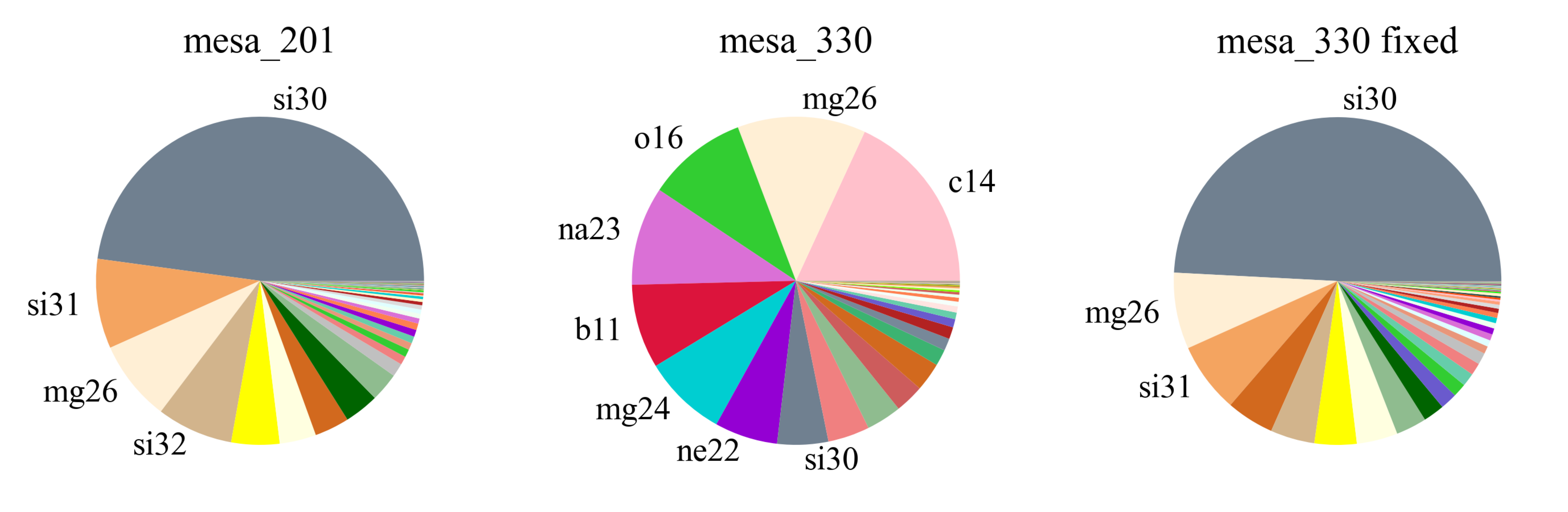}
\vspace*{0.3cm}
\caption{Compositions in nuclear statistical equilibrium for nuclear reaction networks with different sizes. Left panel: equilibrium composition in the case of a 201-isotope network for a \textsc{bbq} run where $T=7\times 10^{9} \K$, $\rho =3\times 10^{8} \g \cm ^{-3}$ (for 80 and 833-3335-isotope networks the final composition is similar). Middle panel: equilibrium composition for a \textsc{bbq} run with the same initial parameters except the nuclear reaction network that contains 330 isotopes (for the 495 isotope network the final composition is similar). Right panel: same as middle panel after fixing the bug in the nuclear reaction networks of \textsc{mesa}.}
\label{fig:CompositionsMESAbug}
\end{center}
\end{figure*}

To find the minimum size of the nuclear reaction network required to give accurate enough values of the composition and electron fraction, and identify the most suitable network for generating the NNNs' training sets, we examined the compositions from \textsc{bbq} runs using networks of varying sizes: 22, 55, 80, 151, 201, 330, 495, 833, 1508, and 3335 isotopes at different timesteps. While looking at large timescales, we found that the equilibrium composition obtained with a 201-isotope network resembles those from larger networks with 833 to 3335 isotopes. The equilibrium composition for \textit{mesa\textunderscore 80} is qualitatively consistent with these larger networks as well. However, two medium-sized networks, with 330 and 495 isotopes, produced notably different equilibrium compositions, as illustrated in Fig \ref{fig:CompositionsMESAbug}. 

From comparing the lists of isotopes between the different nuclear reaction networks we concluded that the presence of tritium in the medium nets (and its absence in the smaller and larger nets) led to this prominent difference in the equilibrium compositions. Specifically, we tracked down the difference to the following reaction
\begin{equation}
\rm n+n+He^{4}+He^{4} \rightarrow h^{3} + Li^{7}, 
\label{eq:reaction}
\end{equation}
which was unexpected given that the probability of a four-particle reaction is typically negligible. We found that the rate of this reaction in \textsc{mesa} is 20-24 orders of magnitude larger than the rate in the literature \citep{MalaneyFowler1989}. 

Exploring the source code, we noticed that \textsc{mesa} miscalculated the rates of the endo-energetic reactions for all reactions involving more than two reactants and/or products. This is because it calculated these reaction rates from detailed balance considerations omitting a phase space factor related to the number of particles involved in the reactions. Even though this bug does not affect most stellar evolution calculations, especially for models that include the standard (commonly used) isotope networks, it has a huge impact on the nucleosynthesis results in high temperatures and densities in networks that include tritium. We fixed the bug in the current \textsc{mesa} version (right panel of Fig. \ref{fig:CompositionsMESAbug}; for more information see this \href{https://github.com/MESAHub/mesa/issues/575}{GitHub issue}).

\newpage

\section{Determining the Nuclear Neural Networks Architecture}
\label{sec:AppendixNNNtries}

\begin{figure*}
\begin{center}
\vspace*{1.0cm}
\includegraphics[width=0.55\textwidth]{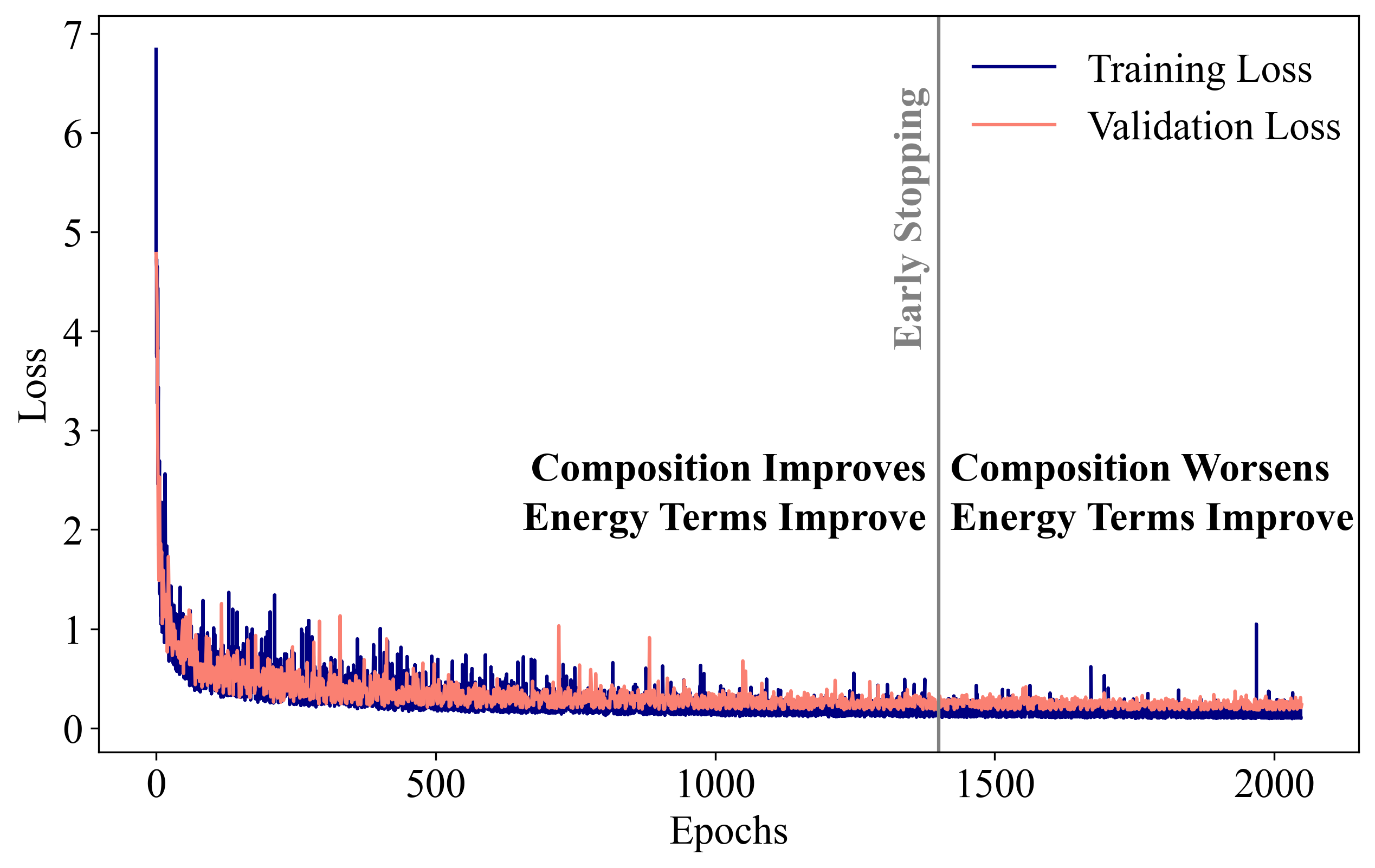}
\vspace*{0.3cm}
\caption{Training (dark blue) and validation (pink) losses vs the number of epochs for nuclear neural networks (NNNs) trained on \textit{mesa\textunderscore datasets} for $dt=0.1 \s$. The gray line marks the point where we applied early stopping since the composition predictions began to worsen. }
\label{fig:LossesVSepoch}
\end{center}
\end{figure*}

To construct the NNN architecture that best predicts the composition and energy terms, we performed an hyper-parameter search. In each case, we monitored the training (dark blue curve in Fig. \ref{fig:LossesVSepoch}) and validation (pink curve in Fig. \ref{fig:LossesVSepoch}) losses, 
and stopped the training (gray line) where the predictions of the compositions began to worsen. We applied this early stopping condition before reaching overfitting in the total loss (i.e., before the total validation loss started increasing) since we prioritize accurate prediction of the electron fraction, which is a key factor influencing SN explosions, over further improving the energy terms.

\begin{figure*}
    \centering
    \subfloat{
        \includegraphics[width=0.485\textwidth]{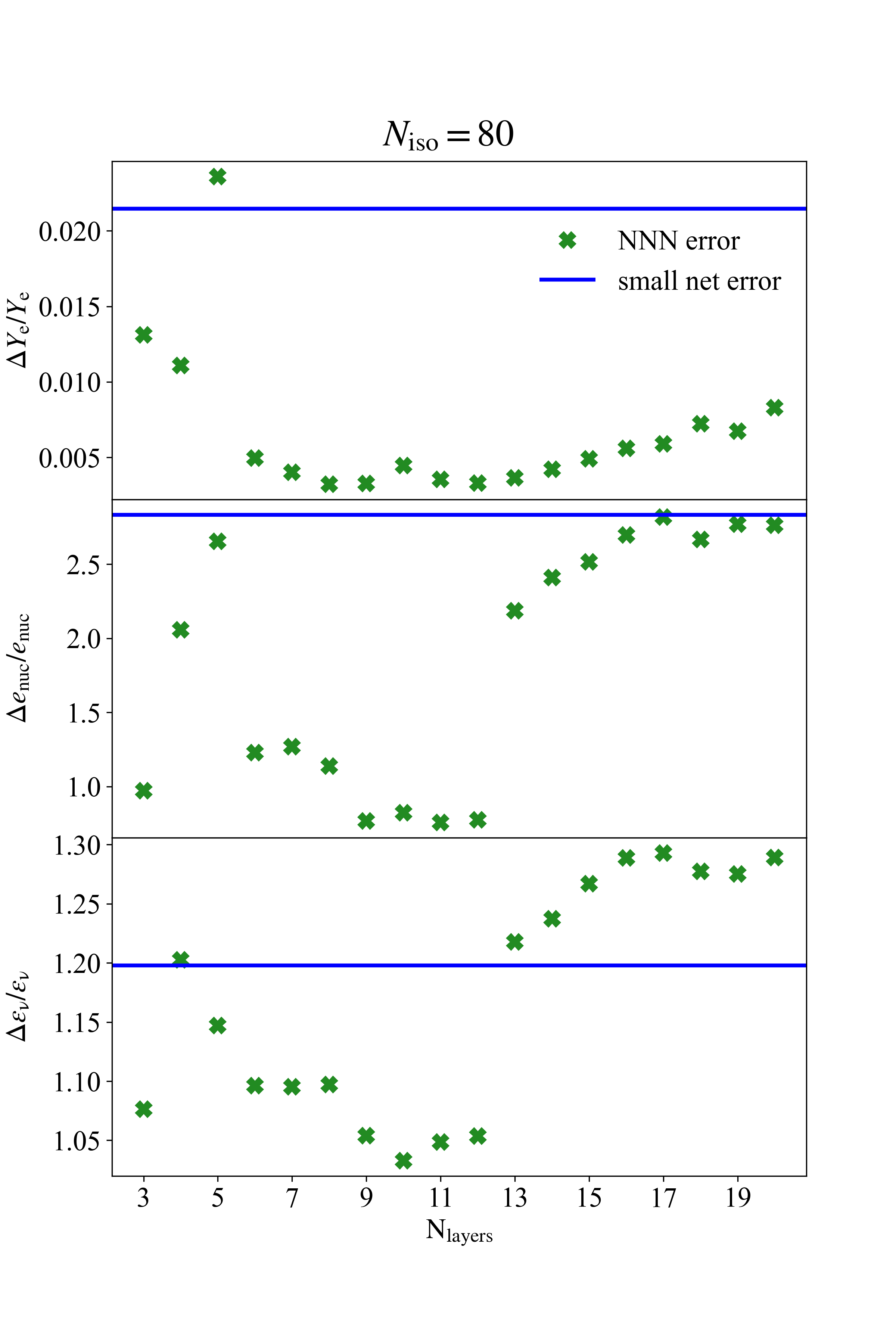}
    }
    \hfill
    \subfloat{
        \includegraphics[width=0.485\textwidth]{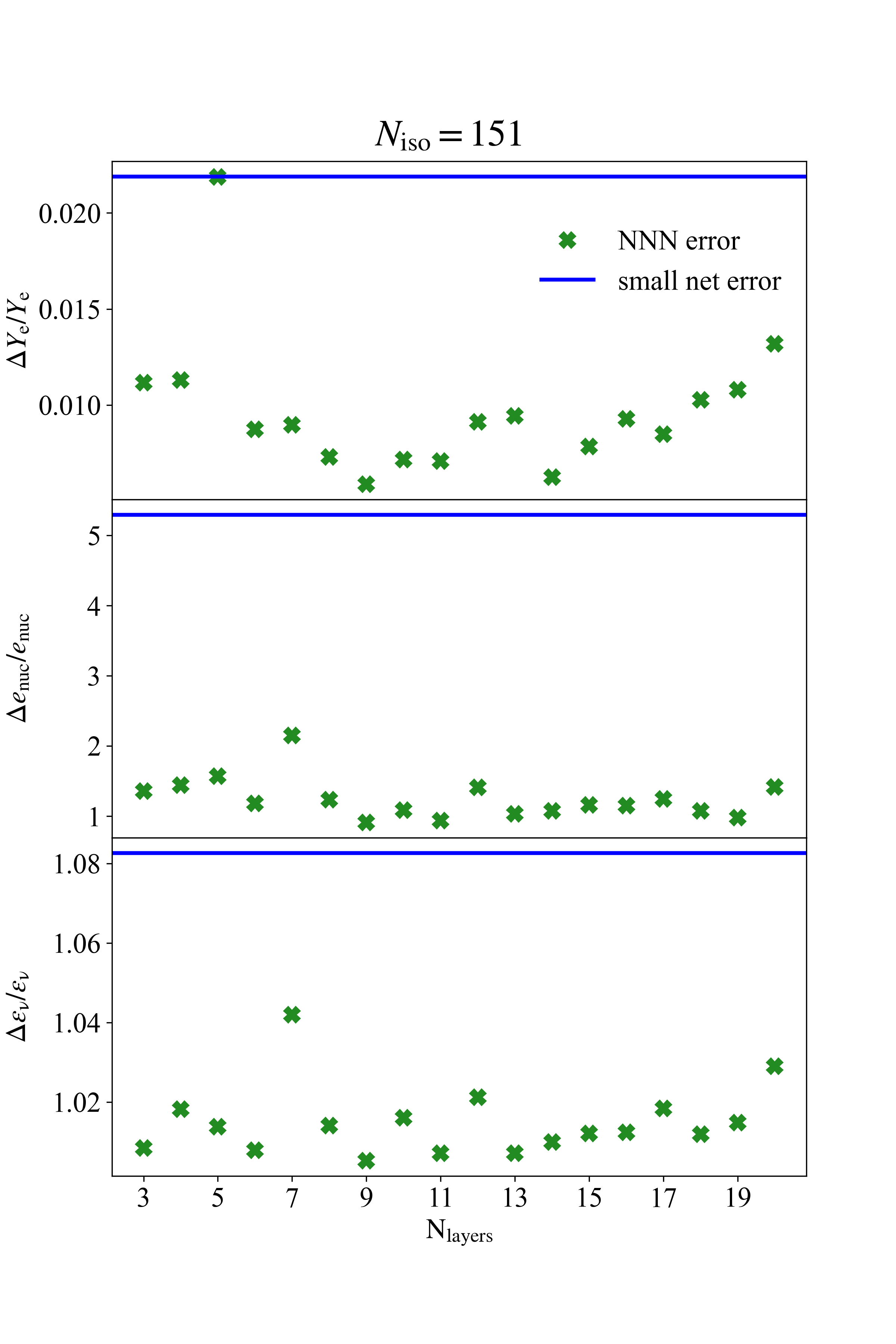}
    }
    \caption{ Comparison between the nuclear neural network (NNN) predictions and the results obtained by using a small nuclear reaction network for different NNN depths. We show the relative error of the electron fraction (top panel), nuclear energy generation term (middle panel) and neutrino-loss term (bottom panel) for NNNs with a different number layers trained on a timestep of $dt = 0.1 \s$. The green X markers are the errors of NNNs trained to predict the composition and energy generation of \textit{mesa\textunderscore 80} (left panel) and \textit{mesa\textunderscore 151} (right panel). The blues lines in both panels mark the average relative error between values found in the large networks and in the small net \textit{approx21\textunderscore cr60\textunderscore plus\textunderscore co56.} }
    \label{fig:LossesDifferentLayers}
\end{figure*}

To find the optimal depth of the NNNs, we trained emulators with a different number of layers on \textsc{bbq} datasets produced at constant temperatures and densities. In Fig.~\ref{fig:LossesDifferentLayers} we show the dependence of the relative error in NNN predictions (green X markers) of the electron fraction (top panel), nuclear energy generation term (middle panel) and neutrino-loss term (bottom panel) for NNNs trained on \textit{mesa\textunderscore 80} datasets (left panel) and \textit{mesa\textunderscore 151} datasets (right panel) at a timestep of $dt=0.1 \s$. The blue lines represent the relative error between the values of the physical quantities computed for \textsc{bbq} runs with \textit{mesa\textunderscore 80} and \textit{mesa\textunderscore 151}, and the values of the same quantities using the default network \textit{approx21\textunderscore cr60\textunderscore plus\textunderscore co56}. Based on Fig.~\ref{fig:LossesDifferentLayers}, we chose the number of layers to be $N_{\rm layers}=12$ for the NNNs we trained on \textit{mesa\textunderscore 80} datasets and
$N_{\rm layers}=9$ for \textit{mesa\textunderscore 151} datasets, as they minimizes the error in both the electron fraction and energy terms. While less computationally efficient, deep nuclear networks can capture more complex patterns across the learning process, which is beneficial for high-dimensional systems as we have here. 

We have tried several other variations on the NNN architectures before converging on the current setup. At first, we trained two separate models to predict the compositions and the normalized energy generation. While this had a small effect on the predictions of the composition, it prolonged by more than a factor of two the GPU time required to predict the energy generation terms, plateauing in a similar level of accuracy as the predictions of our current architecture. Furthermore, potential inconsistencies that could arise from using two separate trained models lead us to abandon that direction. When creating NNNs that predict both the isotopes' abundances and energy generation, we tried to train them on non-normalize energy densities, apply a logarithmic L1 loss function to the normalized and non-normalized neutrino-loss terms (not possible for the energy generation terms since they could result in negative values), give different relative weights to the errors in the compositions and in the energy terms, include  residual connection and add other predicted outputs to steer the composition in the right direction, all resulting in less accurate results.

\end{document}